\documentclass[5p,authoryear]{elsarticle}
\makeatletter 
\def\ps@pprintTitle{%
 \let\@oddhead\@empty
 \let\@evenhead\@empty
 \let\@evenfoot\@oddfoot} 
\makeatother
\usepackage[utf8]{inputenc} 
\usepackage[T1]{fontenc}
\usepackage[english]{babel}
\usepackage[babel=true]{csquotes} 
\usepackage{amsmath} 
\usepackage{mathtools}
\usepackage{amsthm} 
\usepackage{booktabs} 
\usepackage{multirow} 
\usepackage{amssymb} 
\usepackage{amssymb}
\usepackage{svg}
\usepackage{pdfpages}
\usepackage{multirow}
\usepackage{caption}
\usepackage{tabularx}
\usepackage{nicefrac}
\usepackage{lipsum}

\usepackage{placeins}
\usepackage[symbol]{footmisc}

 \usepackage[super]{nth} 

\usepackage{titlesec}
\titleformat{\paragraph}[runin]{\normalfont\normalsize\bfseries}{\theparagraph}{1em}{}
\titlespacing{\paragraph}{0pt}{\parskip}{-\parskip}
\usepackage{amsthm}
\usepackage{algorithm}
\usepackage{algpseudocode}\usepackage[colorlinks=true,linkcolor=blue]{hyperref}%
\usepackage{soul} 
\usepackage{wrapfig}
\usepackage{float}
\usepackage{multicol}

\begin{document}
\sloppy
\begin{frontmatter}

\title{Real-time and On-site Aerodynamics using Stereoscopic PIV and Deep Optical Flow Learning}





\author[1,2,*]{Mohamed Elrefaie}
\author[2]{Steffen Hüttig}
\author[3]{Mariia Gladkova}
\author[2,*]{Timo Gericke}
\author[3]{Daniel Cremers}
\author[1]{and Christian Breitsamter}

\address[1]{Chair of Aerodynamics and Fluid Mechanics, Technical University of Munich, Garching, Germany
}
\address[2]{Volkswagen AG, Wolfsburg, Germany}
\address[3]{Chair of Computer Vision and Artificial Intelligence, Technical University of Munich, Garching, Germany}
\address[*]{Corresponding authors: mohamed.elrefaie@tum.de, timo.gericke@volkswagen.de}

\begin{abstract}
We introduce Recurrent All-Pairs Field Transforms for Stereoscopic Particle Image Velocimetry (RAFT-StereoPIV).  Our approach leverages deep optical flow learning to analyze time-resolved and double-frame particle images from on-site measurements, particularly from the 'Ring of Fire,' as well as from wind tunnel measurements for real-time aerodynamic analysis. 
A multi-fidelity dataset comprising both Reynolds-Averaged Navier-Stokes (RANS) and Direct Numerical Simulation (DNS) was used to train our model. RAFT-StereoPIV outperforms all PIV state-of-the-art deep learning models on benchmark datasets, with a 68\,$\%$ error reduction on the validation dataset, Problem Class 2, and a 47\,$\%$ error reduction on the unseen test dataset, Problem Class 1, demonstrating its robustness and generalizability. In comparison to the most recent works in the 
field of deep learning for PIV, where the main focus was the
methodology development and the application was limited
to either 2D flow cases or simple experimental data, we extend deep learning-based PIV for industrial applications and 3D flow field estimation.  As we apply the trained network to 
three-dimensional highly turbulent PIV data, we are able to obtain flow estimates that maintain spatial resolution of the input image sequence. In contrast, the traditional methods produce the flow field of $\sim$16$\times$ lower resolution.
We believe that this study brings the field of experimental fluid dynamics one step closer to the long-term goal of having experimental measurement systems that can be used for real-time flow field estimation.

\end{abstract}


\end{frontmatter}

\section{Introduction}\label{introduction}
Particle Image Velocimetry (PIV) is a crucial method in experimental fluid mechanics for identifying velocity components in complex flow fields \cite{adrian2005twenty}. The flow medium is initially seeded with tiny particles. Then, a laser sheet is used to illuminate the flow domain so that the particles are visible and a camera records a sequence of images. Eventually, the velocity vectors of the flow can be estimated by analyzing the recorded images.
The most commonly used techniques for flow field estimation from particle images are the cross-correlation and the optical flow methods \cite{westerweel1997fundamentals}. Given a sequence of particle images, the cross-correlation method produces a spatially sparse displacement field. In contrast, optical flow methods can provide a dense displacement field \cite{adrian2005twenty}.

More recently, PIV processing has been approached using deep learning techniques 
\cite{rabault2017performing}, \cite{lee2017piv}, \cite{cai2019dense}, \cite{cai2019particle}, \cite{zhang2020unsupervised},  \cite{lagemann2021unsupervised}, \cite{lagemann2021deep},  \cite{yu2021lightpivnet}, \cite{gao2021robust}, \cite{lagemann2022generalization}, \cite{RAFTyu2023deep}, and \cite{han2023attention} demonstrating remarkable success in planar PIV evaluation. However, the complexity of real-world PIV measurements, such as those encountered in industrial applications, demands a more sophisticated approach than planar PIV can provide. In addition, as experimental setups become more intricate and the volume of analyzed data grows, the aforementioned traditional techniques can become computationally more intensive and time-consuming, thus for volumetric flow field quantification Lagrangian Particle Tracking is widely used.

For instance, in highly complex experiments such as the Ring of Fire (RoF) \cite{huttig2023automotive}, \cite{huettig_RoF}, or wind tunnel measurements \cite{ladwig2023reconstruction} where thousands of images need to be processed to analyze the three-dimensional flow, traditional stereoscopic PIV methods may prove insufficient in terms of efficiency and resolution.



The considerable amount of data generated by the measurement systems necessitates extensive processing times for traditional methods, potentially demanding days or even weeks. To address the inherent limitations of conventional methods in assessing three-dimensional flow, we introduce an automated processing framework for PIV image analysis by incorporating deep learning techniques into the pipeline. This approach ultimately demonstrates the advantages of high automation, resulting in a significant reduction of processing time when compared to non-automated systems. Additionally, this method achieves superior spatial resolution, greatly enhancing the efficiency of the experimental measurements. Additionally, although the training phase for the model's parameters can be resource-intensive, it can be executed offline. Once the deep learning model has been adequately trained, the time required for the estimation process of each image pair is approximately 40 seconds\footnote[1]{For $1024 \times 1024 \, \mathrm{px}^2$ double-frame images on one A100 Nvidia GPU.}. This learning approach facilitates real-time and on-site estimation and, thus, high efficiency while preserving acceptable accuracy levels suitable for practical applications.

To summarize, our contributions are the following:
\begin{itemize}
\item We introduce 
RAFT-StereoPIV to accommodate stereoscopic PIV, showcasing its applicability and potential for a diverse range of real-world scenarios in fluid mechanics (RoF and wind tunnel measurements).
\item  We benchmark our deep learning model on the validation dataset Problem Class 2 \cite{lagemann2021deep} (14.4 thousand fluid flow test cases) and the unseen test dataset Problem Class 1 \cite{cai2019dense} (10 thousand fluid flow test cases). We obtain a significant decrease in error rates compared to the state-of-the-art: a \textbf{68\,$\%$} reduction in error rates on the validation dataset and a \textbf{47\,$\%$} reduction on the unseen test dataset.
\item We conduct extensive hyperparameters tuning to optimize the training of our deep learning model. Furthermore, we introduce a smaller version RAFT-StereoPIV-Small, significantly reducing the original model size from 5.3 million parameters to 2.2 million, which corresponds to a decrease of approximately \textbf{59$\%$}, thereby enhancing efficiency while maintaining robust performance.

\end{itemize}

To the best of the authors' knowledge, our work is the first successful application of deep learning for experimental aerodynamics based on stereoscopic PIV.

The rest of the paper is organized as follows: in Section \ref{sec:sec_Related_Work2}, we briefly describe relevant methods for deep learning-based PIV processing and the measurement systems.  We provide an overview of our approach for flow field estimation using deep learning for on-site and real-time car aerodynamics in Section \ref{sec:sec3_RAFT-StereoPIV}. Finally, we cover the implementation details and present the results on both synthetic data and real-world measurements (Section \ref{sec:sec4_Evaluation}). Limitations and future work are discussed in Section \ref{sec:sec5_FutureWork}.


\section{Related Work}
\label{sec:sec_Related_Work2}
\subsection{Deep Learning for PIV}
Analysis of apparent motion from images by means of neural networks has been extensively researched in the computer vision community and pioneered in~\cite{dosovitskiy2015flownet}. 
Building on such foundational contributions, the recent works of \cite{cai2019dense} and \cite{cai2019particle} have inspired a number of follow-up works on applying deep learning methods to PIV data. The authors developed supervised deep learning models based on convolutional neural networks (CNNs) for estimating dense motion that achieved performance comparable to the state-of-the-art methods for PIV processing at that time. One of the important contributions is the synthetic dataset of fluid flow images, which contains a large number of particle images and the corresponding ground-truth fluid motions.
The dataset provided by \cite{cai2019dense}, however, does not allow for generalization to complex flow fields due to the large gap between synthetic and real-world experimental data.
Motivated by the limitations of the aforementioned dataset, \cite{zhang2020unsupervised} and \cite{lagemann2021unsupervised} proposed unsupervised deep learning models for PIV, where the loss function is composed of different loss terms: photometric loss between two consecutive image frames, consistency loss in bidirectional flow estimates, and spatial smoothness. Their models produce results that are competitive with both traditional PIV approaches and currently used supervised learning-based methods and even surpass them in some challenging flow situations. However, the unsupervised models could not achieve the same accuracy as those of the supervised ones.
A data-driven but slightly different approach was done in a recent article from \cite{wang2022densePINN}, where a physics-informed neural network was used to reconstruct dense velocity fields from sparse experimental PIV or Particle Tracking Velocimetry (PTV) data. The potential of such physics-informed neural networks is that they are not only data-driven but also incorporate physical laws in the training. 

Drawing inspiration from the recent optical flow architecture known as recurrent all-pairs field transforms (RAFT) \cite{RAFT}, numerous studies have built upon this network \cite{lagemann2021deep}, \cite{yu2021lightpivnet}, \cite{lagemann2022generalization}, \cite{RAFTyu2023deep}, and \cite{han2023attention}. They achieved remarkable results in the field of PIV for fluid dynamics, thereby highlighting its significant impact and widespread adoption within the research community. The exceptional performance of such optical flow architecture can be attributed to an extensive training dataset, substantial computational power, and considerable model size, emphasizing accuracy over model efficiency and computational performance. Consequently, training such a model presents challenges. Utilizing four Nvidia Titan X GPUs, the training would typically require approximately 12 days.

In two related papers, \cite{liang2021deepptv} and \cite{liang2023recurrent} present novel deep learning approaches to PTV for efficient and accurate flow field measurements. DeepPTV, the first method, uses a deep neural network to learn complex fluid flow motion from consecutive particle sets, aggregating local spatial geometry information from neighboring particles. The second method, GotFlow3D, introduces a recurrent graph optimal transport-based deep neural network for 3D flow motion learning from double-frame particle sets. This approach constructs two graphs in geometric and feature space, fusing intrinsic and extrinsic features learned from a graph neural network to iteratively and adaptively guide recurrent flow learning. Both methods exhibit state-of-the-art performance in accuracy, robustness, and generalization, offering deeper insights into the complex dynamics of various physical and biological systems.

Despite the advancements in PTV and deep learning for 3D flow field measurements, there remains a gap in the literature regarding flow estimation based on stereoscopic PIV data using deep learning techniques. In addition, much of the existing research has been focused on methodology development and simple test cases rather than addressing the challenges posed by real-world experimental data, which hold significant relevance for industrial applications. In contrast, our work leverages deep learning for stereoscopic PIV data analysis to provide practical solutions for complex industrial flow scenarios.

\subsection{Ring of Fire}
Understanding the aerodynamics is crucial for many aspects of car design development. This includes the drag analysis, rain, dust, and soiling simulation, aeroacoustics, and the validation of Computational Fluid Dynamics (CFD) methods. The majority of experimental research in car aerodynamics is carried out in wind tunnels using pressure probes, hot-wire anemometers, tufts, or through optical metrology methods such as PIV, PTV, or 'Shake-The-Box' \cite{schanz2016shake}. Due to the limitations imposed by the size of the wind tunnels and the measurement techniques employed, researchers often simplify the problem by resorting to stationary scaled models. Therefore, wind tunnel experiments cannot reproduce the exact aerodynamics encountered during real driving conditions like platooning, cornering, and drafting.

Ring of Fire (RoF)  
\cite{sciacchitano2015quantitative}, \cite{terra2017aerodynamic}, 
\cite{terra2018drag}, \cite{spoelstra2018assessment}, 
\cite{spoelstra2018ring},
\cite{spoelstra2019site},  
\cite{spoelstra2020drafting}, \cite{spoelstra2021uncertainty}, \cite{huttig2023automotive}, \cite{huettig_RoF}, \cite{spoelstra2023site} is a novel quantitative measurement technique that can provide insights into the flow topology of full-scale, transiting objects and determine their aerodynamic drag. It is a non-intrusive, large-scale, stereoscopic PIV measurement technique that can be implemented for a full three-dimensional flow field description. 
The test object passes through the measurement section and the flow field is captured in the laser light section \cite{spoelstra2019site}.  In addition to visualizing the flow field, the drag coefficient can be determined by evaluating the momentum loss in a control volume \cite{terra2017aerodynamic}.
For better flow field visualization and accurate drag analysis, a dense and instantaneous velocity field with high resolution as well as resolved fluctuations and turbulent structures is essential \cite{spoelstra2021uncertainty}.  
Having high spatial resolution of the flow field can assist the engineers in better understanding and evaluating the unsteady flow phenomena around the transiting object.

In collaboration with Volkswagen AG \cite{huttig2023automotive}, \cite{huettig_RoF} conducted the first RoF experiment for the automotive industry. The objective of this research was to assess the applicability and feasibility of RoF
system for cars, quantify the flow in the wake of different models and configurations, and additionally determine the drag and its uncertainty through the control volume
approach.

One of the main motivations for the RoF measurement system is to have an on-site flow field estimation.  Moreover, it has the potential to measure dynamic motions without the need for complex moving ground simulations in a wind tunnel or CFD \cite{huttig2023automotive}, \cite{huettig_RoF}. The implementation of such measurement technique can be done under real driving conditions with tire and ground effects and independent of the vehicle. 
By using high-speed cameras with a short acquisition time of 6000 frames per second (FPS) at resolutions of $1024 \times 1024 \, \mathrm{px}^2$, 
high spatial and temporal resolution can be achieved. 
However, the resultant vector field is not immediately available after the measurement, and therefore, neither drag estimation nor flow field visualization \cite{spoelstra2019site}, \cite{huttig2023automotive}, \cite{huettig_RoF}. This is due to the fact that preprocessing of the measurement data takes substantial amount of manual work and processing time when using classical methods. In contrast, our method does not need any preprocessing and allows an automated and real-time flow field estimation as the deep learning model can learn different preprocessing operations and filters from the training data.














\section{Approach: Flow Field Estimation Using Stereoscopic PIV and Deep Learning}
\label{sec:sec3_RAFT-StereoPIV}



\begin{figure*}[ht!]
    \centering
    \includegraphics[width=\textwidth, height=0.9\textheight]{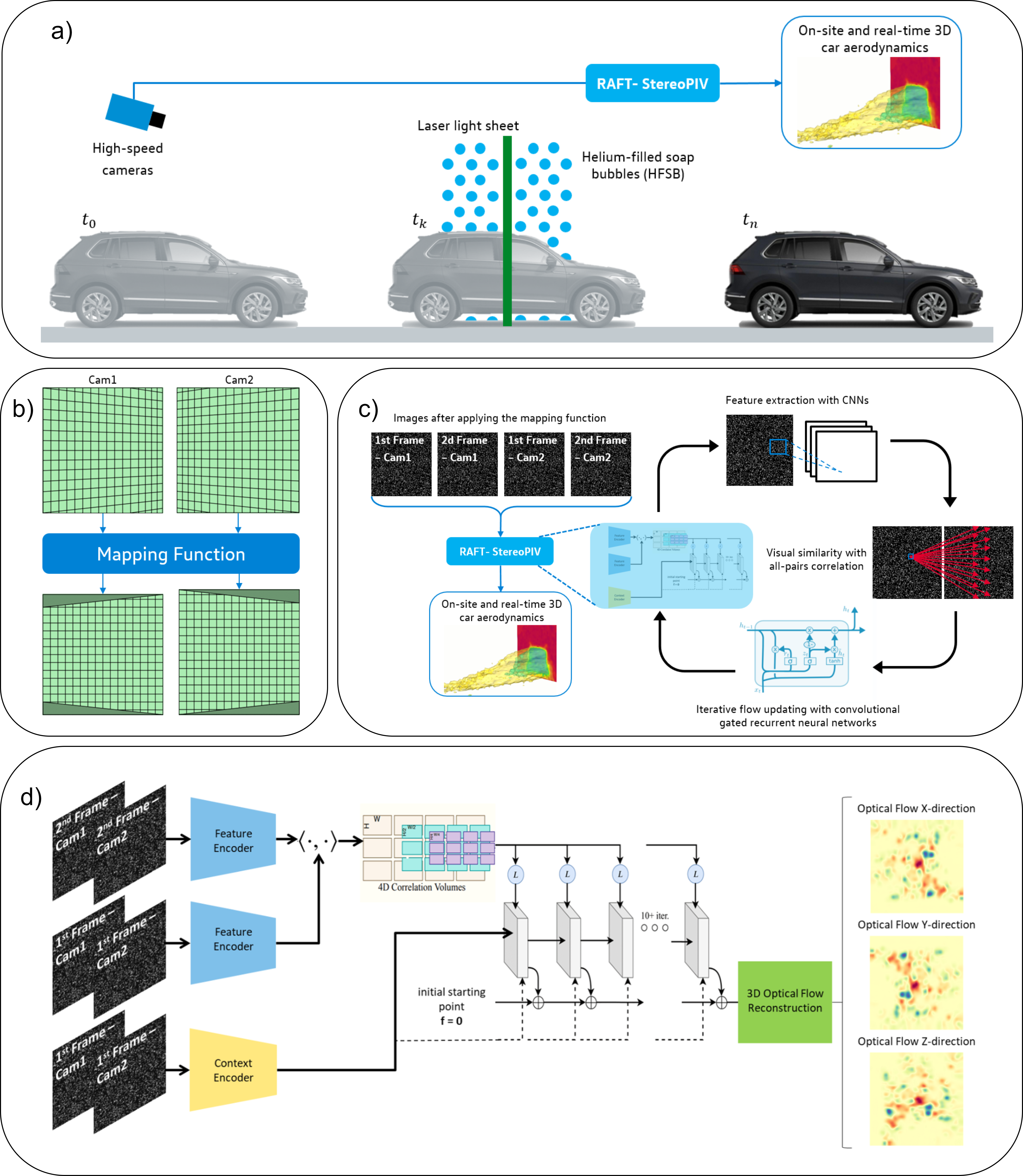}
    \caption{Overview of our approach for flow field estimation using RAFT-StereoPIV. a)  Stereoscopic configuration of Ring of Fire where the car crosses the measurement section and particle images are processed on-site and in real-time by RAFT-StereoPIV to output a 3D vector field. 
     b) The Mapping function used for 3D reconstruction from 2D Stereoscopic PIV Data. 
     c) Inputs to RAFT-StereoPIV and its main operations.
     d) Processing pipeline RAFT-StereoPIV (adapted from \cite{RAFT}), which takes a video image sequence from the stereoscopic configuration and estimates the corresponding optical flows of the in-plane components (in $x$- and $y$-directions) for each camera. The 3D velocity field $(U, V, W)$ can be reconstructed from the 2D optical flow pair as discussed in \ref{sec:3DReconstruction}. 
     }
    \label{fig:overviewRAFTROF}
\end{figure*}

An overview of our approach for flow field estimation using RAFT-StereoPIV is given in Figure  \ref{fig:overviewRAFTROF}. Our method, while demonstrated here in the context of RoF for real-time and on-site car aerodynamics, is adaptable and effective in a wide range of stereoscopic experimental setups.

The RAFT-StereoPIV computational framework comprises four essential steps:
\begin{itemize}
    \item \textit{Feature Extraction and Context
Encoding}: A shared feature encoder is employed to extract per-pixel features from both input images. Utilizing the same architecture as the feature encoder, the context encoder generates a context map for the first image frame. Both encoders use same convolution, where the output size is maintained equal to the input size.
    
    \item \textit{All-pairs
Correlation}: This step involves computing the visual similarity between the two frames by taking the inner product of all pairs of feature vectors.

    \item \textit{Iterative Updates}: Using a convolutional gated recurrent unit (Conv-GRU) multiple iterative updates of the estimated flow field take place. 
    
    \item \textit{3D Reconstruction}: In this step, the three-dimensional velocity field can be reconstructed from a pair of two-dimensional optical flow data.

\end{itemize}

In this approach, features and motion priors are learned by the feature and context encoder, as well as the update operator, all of which are differentiable and integrated into a single end-to-end trainable architecture.

As an input to our framework, we utilize a video
image sequence from the stereoscopic configuration. The deep learning model is then used to estimate the corresponding optical flows of the in-plane components (in $x$- and $y$-directions) for each camera.



To obtain the optical flow out-of-plane (in $z$-direction), we utilize a mapping function for the 3D reconstruction from 2D stereoscopic PIV data based on geometric calibration and digital self-calibration \cite{prasad2000stereoscopic},  \cite{wieneke2005stereo}.

\subsection{Experimental Setup}
\subsubsection{Ring of Fire}
\begin{figure}[ht!]
    \centering
\includegraphics[width=\columnwidth]{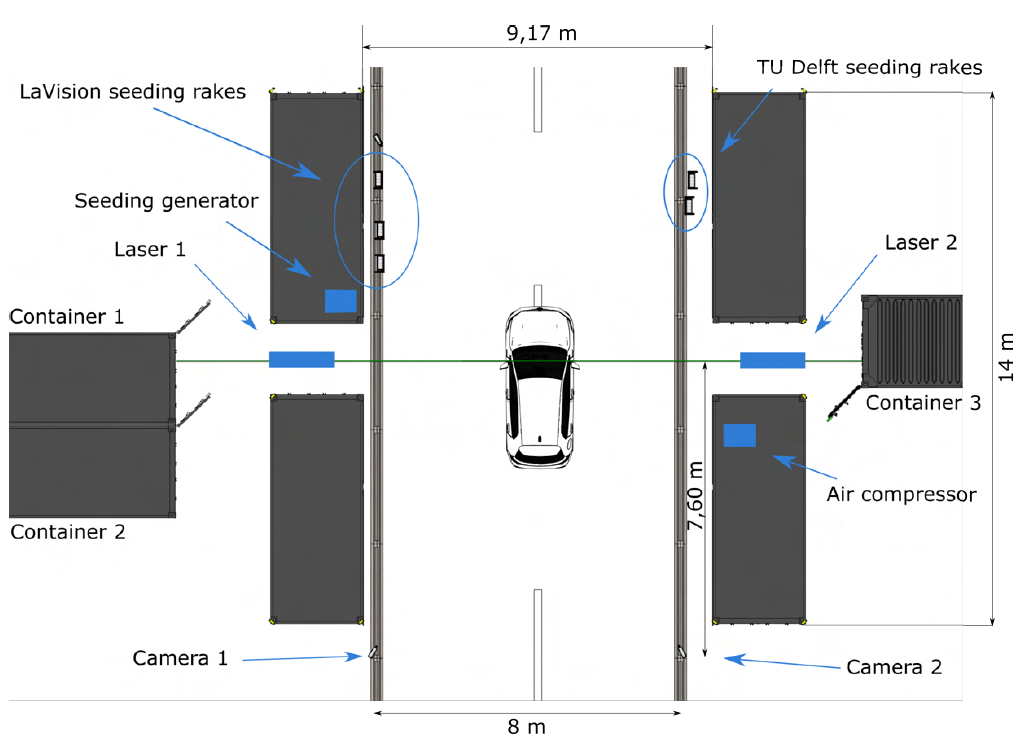} 
\caption{Close up top view of the Ring of Fire experimental setup (\cite{huttig2023automotive}).}
\label{fig:RoF_ExperimentalSetup}
\end{figure}

The RoF experiment \cite{huttig2023automotive}, \cite{huettig_RoF} was conducted outdoors at the Volkswagen AG proving ground in Germany with a tunnel structure to keep the Helium-Filled Soap Bubbles (HFSB) inside the measurement domain and generate a sufficient seeding density. A general overview of the setup is shown in Figure \ref{fig:RoF_ExperimentalSetup}. 
The test object is a Volkswagen up! with a frontal surface area of 2.07 $\mathrm{m}^2$, resulting in a tunnel blockage ratio of 4.5$\%$. The car's body was coated with a matte black foil to reduce light reflections. For the wake analysis, two Photron SA-X2 cameras were set up at the tunnel's entrance, positioned 7.6 m away from the light sheet. 
The cameras have a sensor size of $1024 \times 1024 \, \mathrm{px}^2$, maximum acquisition frequency of 10 $\mathrm{kHz},$ 12-bit digital output, and 20 $\mu \mathrm{m}$ pixel pitch. They were equipped with Scheimpflug-Adaptors v3 from LaVision and Nikon Nikkor lenses with a focal length of $\mathrm{f}=60 \mathrm{~mm}$ and the lens aperture was set to $\mathrm{f}_{\#}=4.0$. The light sheet was about 4 cm thick, created using two high-speed lasers, Quantronix Hawk Pro and Quantronix Hawk II, and relevant light sheet optics; these were positioned on both sides of the road in the middle of the tunnel. Measurements were acquired at a rate of 6 kHz resulting in 6,000 images
for each run.

HFSB were released into the tunnel via fourteen 20-nozzle LaVision seeding arrays and a 204-nozzle-generator seeding rake from TU Delft, reaching a total production of up to 18 million bubbles per second. To allow for a high concentration of particles in the test section, two curtains were set up at the tunnel's entrance and exit, and one was closed during seeding, depending on the flow conditions. The experiments were conducted during the night when the wind was minimal. Each run recorded images covering a 33.33 $m$ distance, about 10 $m$ ahead and 20 $m$ behind the car. The car's cruise control was set to 33.33 $m/s$ when crossing the laser sheet, yielding a Reynolds number of $3.30 \times$$10^{6}$ based on the car's height. 


\subsubsection{Wind Tunnel Measurements of Ahmed Vehicle Model}
\begin{figure*}[htbp]
    \centering
\includegraphics[width=0.95\textwidth]{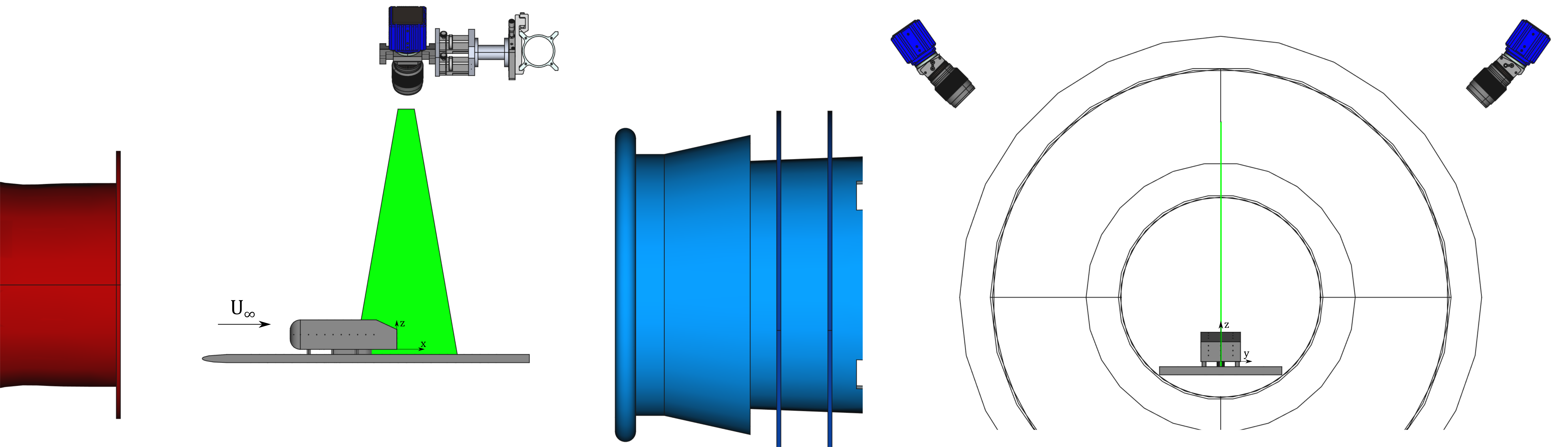}
    \caption{Stereoscopic PIV setup of the 25$\%$-scaled Ahmed body.}
    \label{fig:AhmedBody_Setup}
\end{figure*}
The experiments were conducted in the closed-loop wind tunnel at Bochum University of Applied Sciences \cite{ladwig2023reconstruction} and \cite{gericke2023reconstruction}. This atmospheric tunnel, designed as a Göttingen-type, features a nozzle exit with a diameter of 0.5 meters. The semi-closed test section has a length of 1.2 meters. The mesh screen of the wind tunnel was removed for this measurement campaign to prevent the accumulation of HFSB. The turbulence intensity was recorded at or below 3.5$\%$ for this setup. A flow rectifier with a 20 mm mesh size was placed 100 mm upstream from the mesh's intended position in the wind tunnel's settling chamber. A flat plate with an elliptical nose, a length of 800 mm, and a width of 300 mm is centered 200 mm after the nozzle’s exit. On top of the flat plate, a 25$\%$-scaled model of the Ahmed body reference \cite{ahmed1984some} with a 25$^{\circ}$ slant angle was mounted 215 mm behind the plate's leading edge. The dimensions of the Ahmed body are $261  \times  97  \times  72$ mm in length, width and height, respectively. To reduce noise from laser light reflection, all elements, including the plate and the model, were painted black. Under the Ahmed body, a NACA0020 profile was incorporated to cover the plastic tubes connecting the pressure taps and Scanivalve modules. 
The stereoscopic PIV setup consists of two Imager CX12 Cameras from LaVision (4080 $\times$ 2984 px\textsuperscript{2}; 2.7~\textmu m pixel size, 12-bit) equipped with Scheimpflug adaptors v3 and Nikon Nikkor f=60 mm with an aperture of f\textsubscript{\#}=2.8. Illumination of the DEHS particles (mean diameter $\approx$ 1 \textmu m) was realized by a PIV-Laser Quantel Evergreen 200 (repetition rate 15 Hz, 200 mJ) and corresponding light sheet optics. Control and triggering of the cameras and the laser was performed via DaVis 10.2 and a programmable timing unit.
The experimental setup is shown in Figure \ref{fig:AhmedBody_Setup}.

\subsection{3D Reconstruction from 2D Stereoscopic PIV Data}
\label{sec:3DReconstruction}
\begin{figure}[h!]
    \centering
\includegraphics[width=\columnwidth]{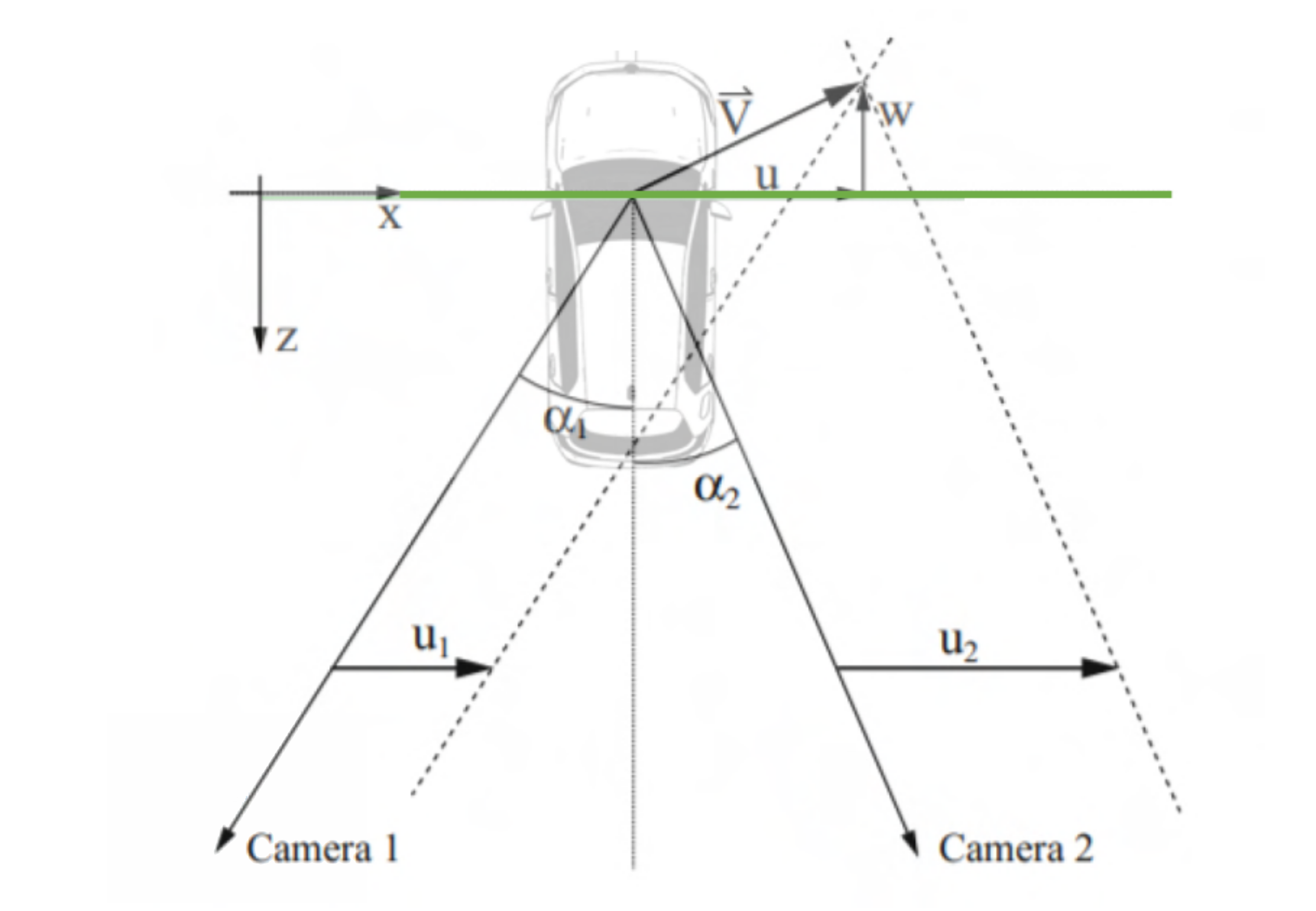} 
\caption{Out-of-plane velocity component reconstruction.}
\label{fig:RoF_angles}
\end{figure}
Since the deep learning model was trained only on 2D-planar particle images and the corresponding optical flows are in-plane components (in $x$- and $y$-directions), optical flow out-of-plane (in $z$-direction) can not be directly estimated using the pretrained network. Based on a non-symmetric stereoscopic arrangement, accurately reconstructing the out-of-plane velocity component is possible because the identical object is captured by two cameras simultaneously, but from different perspectives \cite{prasad2000stereoscopic}.
Consequently, true 3D velocity field $(U, V, W)$ can be reconstructed from a pair of 2D displacements $(\Delta x_1, \Delta y_1)$ and  $(\Delta x_2, \Delta y_2)$ as seen from left and right camera respectively by knowing the precise position of the cameras with respect to the measurement plane.
The 3D flow field estimation is only possible within the area covered by both cameras from the stereoscopic configuration. However, the overlap area is a trapezoidal region of the light sheet due to perspective distortion. Accordingly, the overlap area should be maximized by accurate alignment and calibration \cite{willert1997stereoscopic}, \cite{wieneke2005stereo}.

A mapping function is, therefore, required to estimate the projected displacement fields. This mapping function is obtained from a calibration process that defines an accurate relation between the object and the image space. First, geometric calibration of the system is performed using the pinhole model from \cite{prasad2000stereoscopic}. A calibration plate was aligned with the laser pointing axis and perpendicular to the driving direction of the car. For this, a calibration grid with known dimensions, which consists of markings distributed over two levels of depth, is placed inside the light sheet. The pinhole approach approximates the camera’s positions with respect to each point on the calibration grid by means of triangulation. For each camera, the second-order mapping function, as employed by \cite{willert1997stereoscopic}, consists of estimating twelve coefficients. A minimum of six calibration points is necessary to determine the coefficients based on the least-squares method. \cite{wieneke2005stereo} further proposes to perform a self-calibration procedure on 5-50 image pairs showing particles in motion. A disparity map between the image planes is then created, using the ensemble-averaged cross-correlation. The calibration error can be reduced even more by shrinking and deforming the interrogation windows of the disparity map, based on the previous self-calibration procedure.

The in-plane velocity components measured by the left camera can be described by \cite{prasad2000stereoscopic}:
\begin{equation}
u_1=\frac{\Delta x_1}{M \cdot \Delta t} \quad \text {and} \quad v_1=\frac{\Delta y_1}{M \cdot \Delta t},
\label{eq:DxDt}
\end{equation}
with $\Delta x_1$ and $\Delta y_1$ being the displacements as seen from the left camera, $M$ being the magnification factor, and $\Delta t$ serving as the change in time.
Similarly, the in-plane velocity components $u_2$ and $v_2$ measured by the right camera can be calculated using Eq. \ref{eq:DxDt}.
The velocity components can be reconstructed according to the following equations \cite{willert1997stereoscopic},  \cite{prasad2000stereoscopic}:
\begin{equation}
U=\frac{u_1 \tan \alpha_2+u_2 \tan \alpha_1}{\tan \alpha_2+\tan \alpha_1},
\end{equation}

\begin{equation}
V=\frac{v_1 \tan \beta_2+v_2 \tan \beta_1}{\tan \beta_2+\tan \beta_1},
\end{equation}

\begin{equation}
W=\frac{u_1-u_2}{\tan \alpha_2+\tan \alpha_1},
\end{equation}

where $\alpha_{1/2}$, given in Figure \ref{fig:RoF_angles}, represent 
the angles between cameras 1 and 2 and the z-axis in XZ-plane respectively, $\beta_{1/2}$ are the angles between the cameras and z-axis in the YZ-plane side view.

\subsection{The Deep Learning Model}
Our network architecture is based on RAFT \cite{RAFT}, 
\cite{lagemann2021deep}, \cite{yu2021lightpivnet}, \cite{lagemann2022generalization}, \cite{RAFTyu2023deep}, and \cite{han2023attention}.

A shared convolution-based feature encoder network is applied to extract features from grayscale particle image-pair $I^t$ and $I^{t + \Delta t }$. The feature encoder is fed with low-resolution images ($32 \times 32 \, \mathrm{px}^2$) and maps them to a dense feature map with the same resolution:
\begin{equation}
\mathbf{w}^{(I^t)} = g_\theta\left(I^t\right), \quad \mathbf{w}^{(I^{t + \Delta t })} = g_\theta\left(I^{t + \Delta t }\right) : 
\end{equation}
$$
 \mathbb{R}^{32 \times 32} \rightarrow \mathbb{R}^{32 \times 32 \times 256}
 $$

\begin{figure}[b!]
    \centering
\includegraphics[width=0.8\columnwidth]{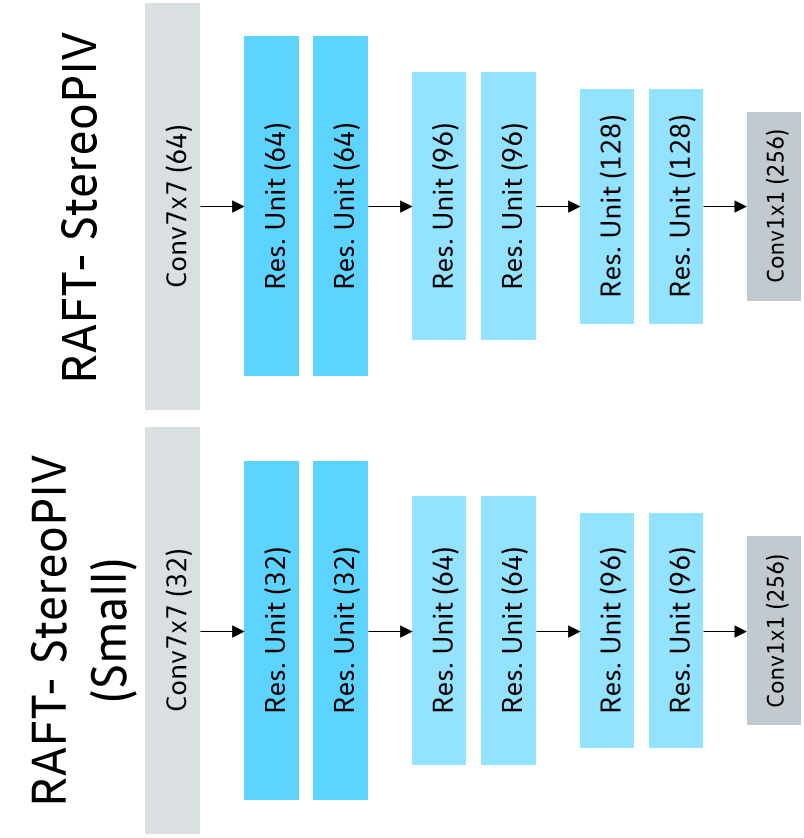}
\caption{Feature and context encoder architectures of  RAFT-StereoPIV and RAFT-StereoPIV-Small models.}
\label{fig:Encoder}
\end{figure}

Additionally, a context network, which has the same architecture as the feature encoder network is applied to the first input image $I^t$ to obtain a context feature map that will be fed later to an update block. Both the feature and context encoders in Figure \ref{fig:Encoder} consist of two convolutional layers and six residual layers, using ReLU as an activation function and instance normalization for normalization. We also introduce RAFT-StereoPIV-Small variant, a smaller model, where the residual units are replaced with bottleneck residual units, which are designed to reduce computational complexity and the number of
parameters. This alternative model is considerably more compact, with each of its feature and context encoders containing only about 80,000 parameters. This presents a significant contrast to the larger variant of the model, where each encoder has approximately 1 million parameters, demonstrating the efficiency and reduced computational burden of the RAFT-StereoPIV-Small model.

The dot product is then used as a measure of visual similarity between all pairs of the feature maps $\mathbf{w}^{(I^t)}$ and $\mathbf{w}^{(I^{t + \Delta t })}$ obtained from the feature encoder. This results in a full correlation volume between all pairs, which can be described as a single matrix multiplication:
$$
\mathbf{C}\left(\mathbf{w}^{(I^t)}, \mathbf{w}^{(I^{t + \Delta t })}\right) \in \mathbb{R}^{32 \times 32 \times 32 \times 32}
$$
\begin{equation}
C_{i j k l}=\sum_h \mathbf{w}^{(I^t)}_{i j h} \cdot \mathbf{w}^{(I^{t + \Delta t })}_{k l h}
\end{equation}

\begin{figure}[htp!]
    \centering
\includegraphics[width=\columnwidth]{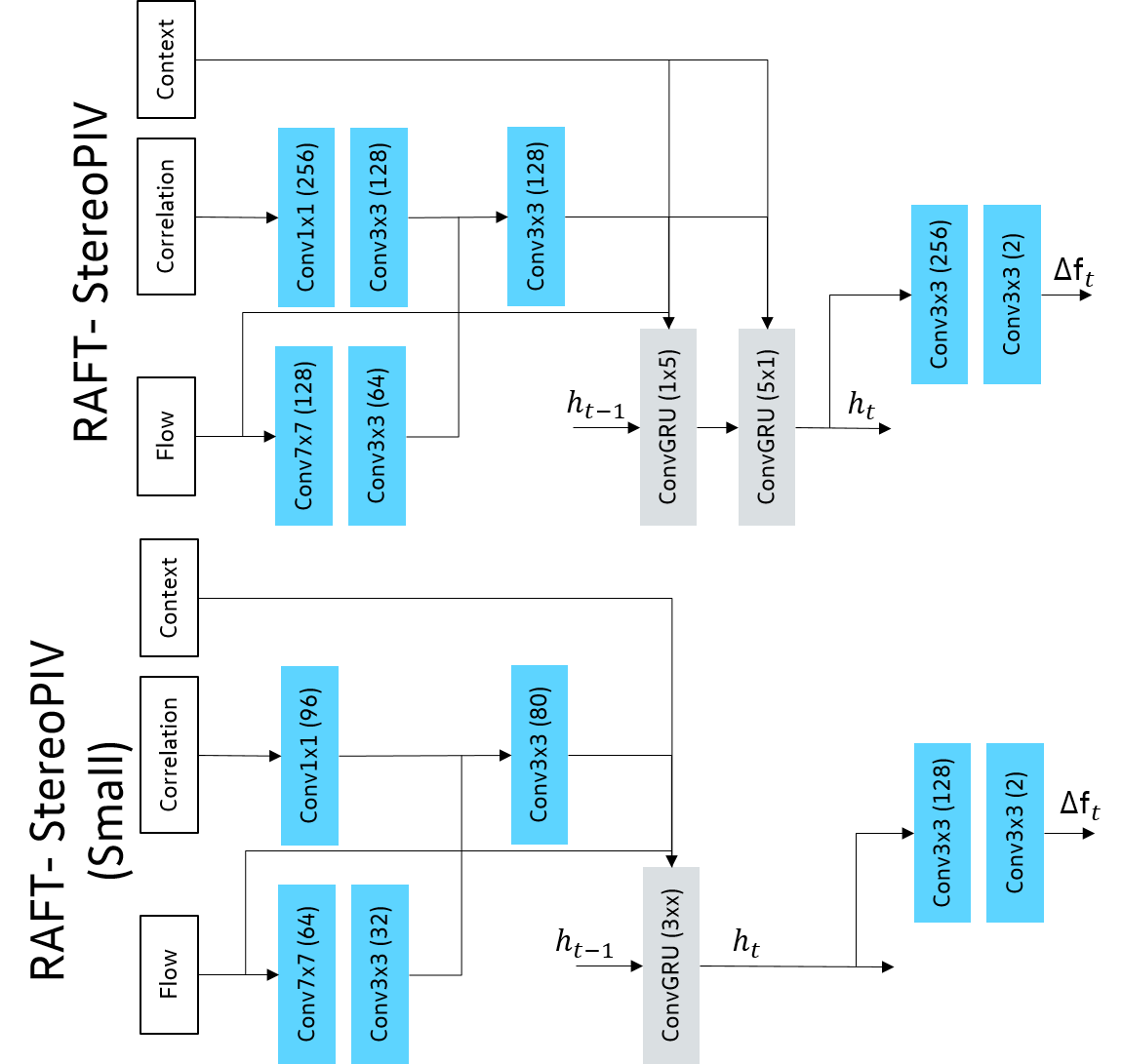}
\caption{Update block architecture of RAFT-StereoPIV and RAFT-StereoPIV-Small models. 
}
\label{fig:UpdateBlock}
\end{figure}

After computing the correlation volume, a 4-layer correlation pyramid $\left\{\mathbf{C}_1, \mathbf{C}_2, \mathbf{C}_3, \mathbf{C}_4\right\}$ is constructed by average pooling the last two dimensions of the correlation volume with kernel sizes 1, 2, 4, and 8 and equivalent stride. The pooling operation acts as a feature selection step. That is, it selects the best features that describe the particle images. 
Furthermore, a correlation feature map is obtained using a lookup operation by indexing from the correlation pyramid. This indexing is performed within a local neighborhood $\mathcal{N}\left(\mathbf{x}^{\prime}\right)_r$, given by Eq. \ref{eq:CorrelationLookup}, where each pixel $\mathbf{x}=(u, v)$ in $I^t$ is mapped to $\mathbf{x}^{\prime}=\left(u+f^1(u), v+f^2(v)\right)$ with $f^1$ and $f^2$ the current estimated optical flow. 
\begin{equation}        \mathcal{N}\left(\mathbf{x}^{\prime}\right)_r = \left\{\mathbf{x}^{\prime}+\mathbf{d x} \mid \mathbf{d x} \in \mathbb{Z}^2, \|\mathbf{d} \mathbf{x}\|_\infty \leq r\right\}, 
    \label{eq:CorrelationLookup}
\end{equation}
$$
\text{where } r=4\,\text{px}
$$
The grid $\mathcal{N}\left(\mathbf{x}^{\prime} / 2^k\right)_r$ is used to index the correlation volume at level $\mathbf{C}_k$. 
In the last step, an update operator is used, which is intended to resemble the steps of a first-order descent optimization algorithm.
As a result, tied weights across depth with bounded activation functions are used to enhance training stability and encourage convergence to a fixed point.
An integral part of the iterative update operator is a gated activation unit based on the GRU cell, with convolutions used in place of fully connected layers. The architectures of the iterative update for the large and the small models are given in Figure \ref{fig:UpdateBlock}.  The larger variant has approximately 3.1 million parameters, while the smaller model's update block contains about 2 million parameters.

The applied Conv-GRU consists of two main gates: the update and the reset gate. It takes the flow, the correlation, and the hidden state $h_{t-1}$ and yields a new hidden state $h_t$, which is then passed through two
convolutional layers and finally outputs the flow update $\Delta \mathbf{f}$. The output of the update gate $z_t$ for the time step $t$ is calculated as:
\begin{equation}
z_t=\sigma\left(\operatorname{Conv}_{3 \times 3}\left(\left[h_{t-1}, x_t\right], W_z\right)\right),
\end{equation}
where $x_t$ denotes the concatenation of flow, correlation, and context features, $h_{t-1}$ the hidden state information of the previous units, $\sigma$ the sigmoid activation function, and $W_z$ the weight matrix of the corresponding unit. The update gate gives the model the ability to choose how much data from previous iterations will be included in the predictions made in the future.
The reset gate $r_t$ is the second important part of the Conv-GRU and its output is computed in the same manner as the update gate but with a different weight matrix:
\begin{equation}
r_t=\sigma\left(\operatorname{Conv}_{3 \times 3}\left(\left[h_{t-1}, x_t\right], W_r\right)\right)
\end{equation}
Based on the output of the reset gate $r_t$, a new memory content $\tilde{h_t}$ is computed using the tanh activation function and the weight matrix $W_h$ in order to store the relevant information:
\begin{equation}
\tilde{h_t}=\tanh \left(\operatorname{Conv}_{3 \times 3}\left(\left[r_t \odot h_{t-1}, x_t\right], W_h\right)\right)
\end{equation}
where $\odot$ denotes the Hadamard (element-wise) product operator. 
Finally, the hidden state vector $h_t$ is obtained using the output of the update gate, the memory content, and the hidden state vector at the previous time step:
\begin{equation}
h_t=\left(1-z_t\right) \odot h_{t-1}+z_t \odot \tilde{h_t}
\end{equation}

It is worth mentioning that each local update incorporates knowledge of both small and large displacements since correlation volumes are constructed for all pairs at both high and low resolution. A total number of 12 iterative updates are used for the optical flow estimation.  The hidden	dimension is 128 for RAFT-StereoPIV and 96 for RAFT-StereoPIV-Small. In the RAFT-StereoPIV model, two convolutional GRU update blocks with 1$\times$5 and 5$\times$1 filters are used, whereas RAFT-StereoPIV-Small model utilizes a single GRU with 3$\times$3 filters.

Both models output optical flow of the in-plane directions $x$ and $y$.
The training of the networks is supervised on the $l_1$ loss with exponentially
increasing weights between the predicted $\mathbf{f}_{i}$ and ground truth flow $\mathbf{f}_{g t}$ over the full sequence of predictions, $\left\{\mathbf{f}_1, \ldots, \mathbf{f}_N\right\}
$:
\begin{equation}
        \mathcal{L} = \sum_{i=1}^N \gamma^{N-i}\left\|\mathbf{f}_{g t}-\mathbf{f}_i\right\|_1,
\end{equation}
$$
\text{where } \gamma=0.8 \quad \text{and} \quad N=12
$$

Averaged endpoint error (AEE) is used as an evaluation metric and it represents the average of the L1-norm of the estimated and the ground-truth optical flow over all pixels: 
\begin{equation}
\mathrm{AEE}=\left\|\mathbf{f}_{\mathrm{es}, N}-\mathbf{f}_{\mathrm{gt}}\right\|_1 
\end{equation}

\section{Evaluation}
\label{sec:sec4_Evaluation}

In this section, we will thoroughly assess the performance and characteristics of the proposed method. To this end, we will utilize different datasets, each offering unique challenges and properties, and conduct a series of experiments. 

\subsection{Datasets}
\subsubsection{Open-source Synthetic PIV Datasets}
\begin{figure*}[h!]
    \centering
    \includegraphics[width=0.9\textwidth]{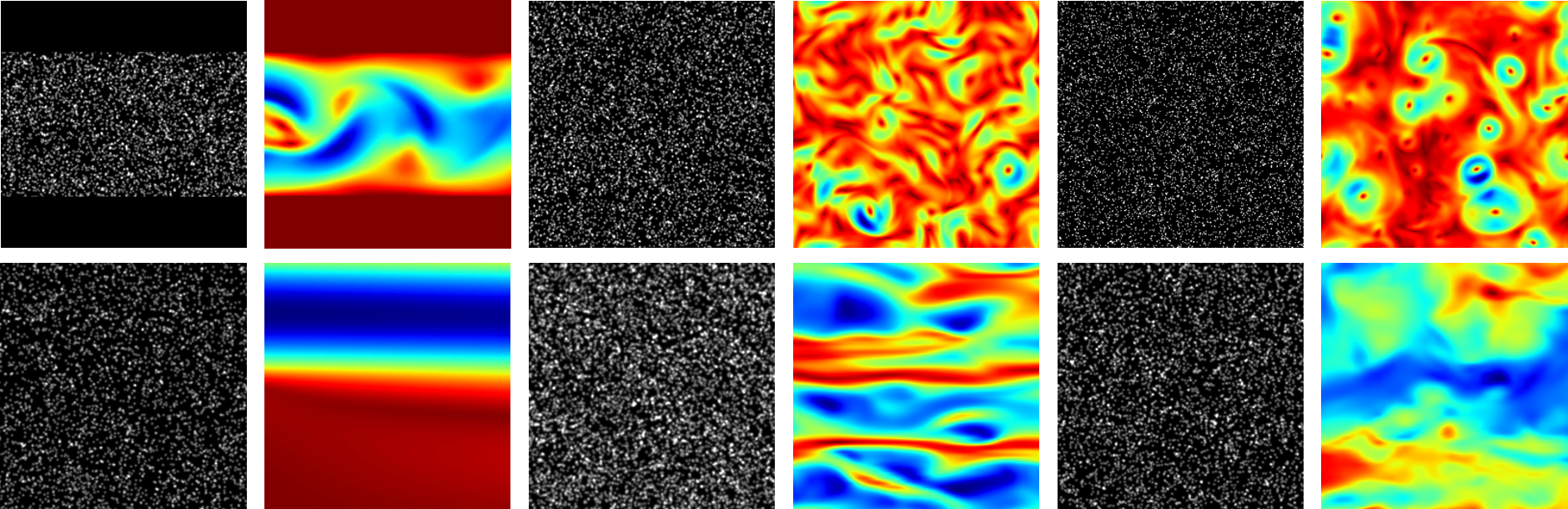}
    \caption{Particle images and the corresponding optical flow from the open-source synthetic PIV dataset Problem Class 2 used to train RAFT-StereoPIV and RAFT-StereoPIV-Small.}
    \label{fig:Synthetic}
\end{figure*}
\paragraph{\textnormal{\textbf{Problem Class 1}}}\mbox{}
\newline
The Problem Class 1 dataset, introduced by \cite{cai2019dense}, consists of diverse fluid motion structures that have been generated using CFD. This dataset includes simulations of various flow cases, such as Cylinder, which represents flow over a circular cylinder, in addition to those sourced from the Johns Hopkins turbulence databases \cite{lee2015direct}. Moreover, the dataset contains cases involving direct numerical simulations (DNS) of turbulence and sea surface flow driven by a Surface Quasi-geostrophic (SQG) model, as originally presented in \cite{carlier2005second} and \cite{resseguier2017geophysical}, respectively.
The dataset is distinguished by its high particle density, particle peak intensity ranging from 200 to 255 counts within an eight-bit grayscale, and maximum particle displacement of $\pm$10\,px. These characteristics produce images that closely resemble those obtained in ideal experimental conditions.


\paragraph{\textnormal{\textbf{Problem Class 2}}}\mbox{}
\newline
The public Problem Class 2 dataset \cite{lagemann2021deep} is based on the 
same ground-truth flow fields of the synthetic dataset of Problem Class 1, however, with reduced particle density, intensity, and overall signal-to-noise ratio (SNR) in order to model image conditions similar to the experimental data. The maximum in-plane displacement was increased from $\pm$\,10\,px to $\pm$\,24\,px. The particle images have a mean SNR value of approximately 3.6, ranging from $[0.1, 70]$ and Gaussian noise with a mean $\mu_n=0.01$ and a variance $\sigma_n=0.075$ based on the normalized pixel values.

Problem Class 2 dataset consists of 15000 particle image-pairs with their corresponding
ground truth flow field in $x-$ and $y-$direction and 
is split into 12000 ($80\,\%$) for training and 3000 ($20\,\%$) for validation. The input data used for training our models RAFT-StereoPIV and RAFT-StereoPIV-Small is smaller chunks from Problem Class 2 of size  $32 \times 32 \, \mathrm{px}^2$ and therefore comprises 100,000 samples for training and 15,000 for validation. Some examples of the particle images and the corresponding optical flow from the open-source synthetic PIV dataset used to
train the deep learning models are shown in Figure \ref{fig:Synthetic}.

A comparison between the characteristics of the two datasets Problem Class 1 and Problem Class 2 is given in Table \ref{tab:problem_classes_1_2}. 
\begin{table}[ht]
\centering
{\scriptsize
\begin{tabular}{|c|c|c|}
\hline
& \textbf{Problem Class 1} & \textbf{Problem Class 2} \\
\hline
\textbf{Particles Diameter} & $2 \mathrm{px} \lesssim d_p \lesssim 6 \mathrm{px}$ & $0.22 \mathrm{px} \leq d_p \lesssim 5 \mathrm{px}$ \\
\hline
\textbf{Particles Density} & $N_{p p p} \gtrsim 0.025$ & $0.005 \leq N_{p p p} \leq 0.1$ \\
\hline
\textbf{Background Noise} & $\mathrm{SNR}>10$ & $1 \leq \mathrm{SNR} \leq 16$ \\
\hline
\textbf{Max Displacement} & $\mathbf{V}_{\max } \leq 10 \mathrm{px}$ & $\mathbf{V}_{\max } \leq 16 \mathrm{px}$ \\
\hline
\textbf{Out-of-Plane Motion} & $w / \Delta z=0$ & $w / \Delta z \leq 0 . \overline{0}$ \\
\hline
\end{tabular} }
\caption{Overview of the public benchmark
datasets and their characteristics.}
\label{tab:problem_classes_1_2}
\end{table} 

We implemented a significant optimization to Problem Class 2 dataset by transitioning the image storage format from \textit{float32} to \textit{uint8}. This modification resulted in a substantial reduction of the dataset's size from 26.52 GB to 8.22 GB, thereby enhancing data efficiency and processing speed.

\subsubsection{Synthetic Dataset of Varying Particle Image Conditions}
To evaluate the performance of RAFT-StereoPIV under varying particle image conditions, we conduct a systematic analysis using synthetic PIV images with a range of parameters. In this study, we focus exclusively on varied image parameters, considering factors such as particle diameter, particle seeding density, particle out-of-plane motion, and the introduction of Gaussian noise to the images. This comprehensive examination enables us to assess the robustness and generalizability of the RAFT-StereoPIV model across various scenarios.

Synthetic images offer considerable advantages, as they allow for precise control over a wide range of image parameters, which is often unattainable in experiments due to inherent uncertainties such as local density and temperature fluctuations, laser and imaging optics-related uncertainties, or pulse-to-pulse variations in laser power and energy distribution. Furthermore, independent manipulation of image parameters enables the examination of their specific effects, which is frequently challenging in measurements, as different parameters may be interdependent or influenced by a common cause. Consequently, the analysis of inference results from these synthetic images yields crucial insights into the most significant image parameters, enhancing our understanding of their impact on the overall system.

\subsubsection{Ring of Fire and Wind Tunnel Experimental Data}
\begin{figure}[h!]
    \centering
\includegraphics[width=\columnwidth]{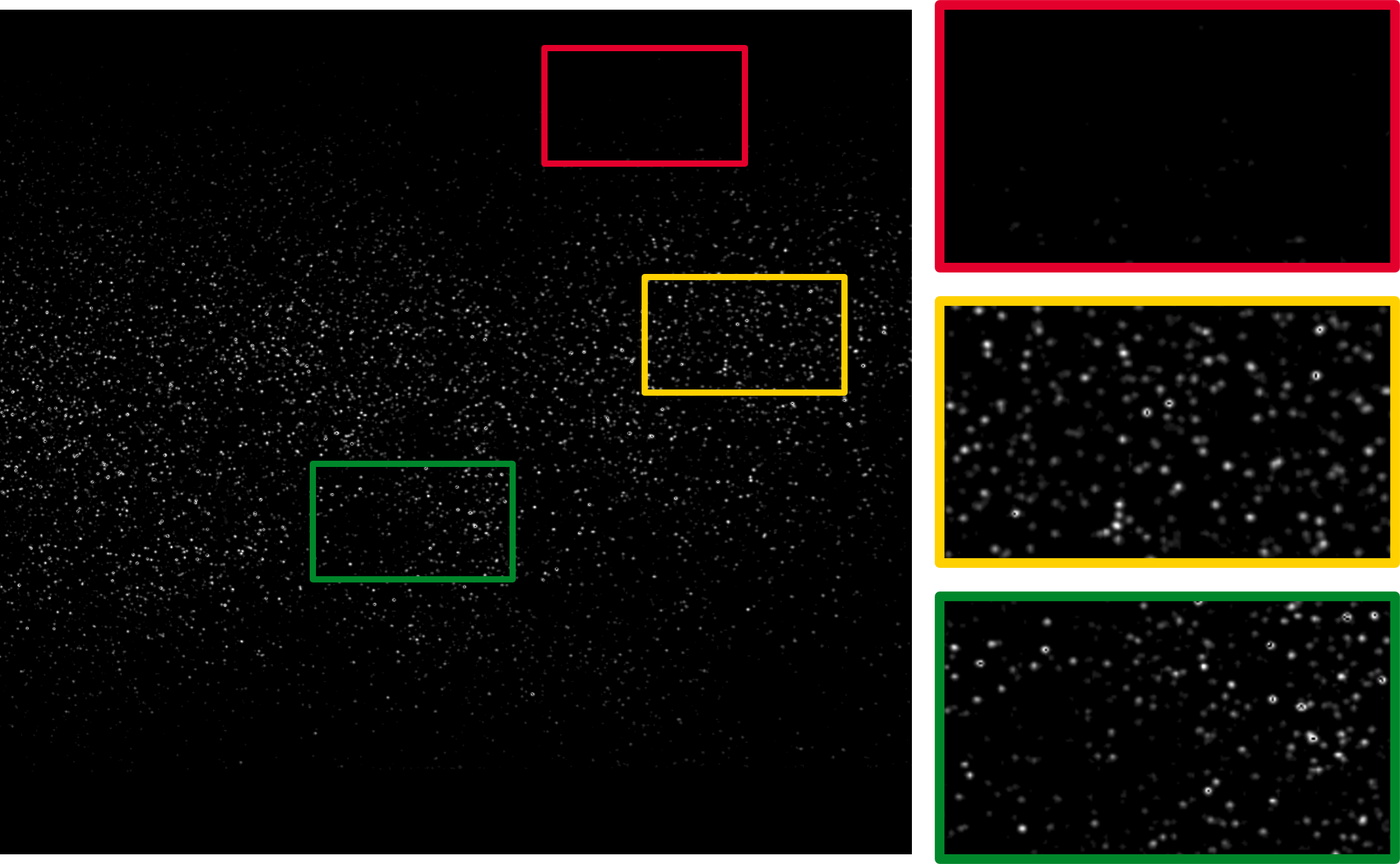} 
\caption{Ring of Fire experimental data after applying the mapping function. Here, we visualize challenging nature of the data, where red zoomed-in region indicates the low seeding density of the particles. The yellow patch shows large variation in the particles' size. The green one shows both low seeding density and variation in the particles' size.}
\label{fig:RoFImage}
\end{figure}
The particle images from RoF and the wind tunnel measurements are imperfect and the flow field estimation is challenging due to several factors: highly turbulent three-dimensional flow within the light sheet results in high uncertainty, non-uniform distribution of seeding particles, and different flow directions with local gradients within and perpendicular to the laser sheet plane as seen in Figure \ref{fig:RoFImage}.
Compared to synthetic images used for training the models (see Figure \ref{fig:Synthetic}), we can see that the experimental data is more challenging and provides larger uncertainties and error sources.


\subsection{Implementation Details}
Our RAFT-StereoPIV and RAFT-StereoPIV-Small models are implemented in the open-source framework Pytorch \cite{pytorch}. In our training, we parallelize across two Quadro GV100 GPUs. RAFT-StereoPIV has approximately 5.3 million trainable parameters. In contrast, its more streamlined counterpart, RAFT-StereoPIV-Small, maintains a compact structure with around 2.2 million trainable parameters. All modules are initialized from scratch with random weights. We follow the same approach as \cite{RAFT} and clip gradients to the range $[-1,1]$.
During training, we use the AdamW optimizer \cite{AdamW} with $\beta_1=0.9$, $\beta_1= 0.999$, and weight decay $\lambda$ = 0.01. The bigger model was trained with a batch size of 196 for 120 epochs, while the smaller one was trained with a batch size of 250 for 60 epochs. During training, the optical flow is initialized to zero in the update block. However, during inference on the RoF images, we apply the warm-start initialization, where the optical flow estimated from the previous pair of frames is used as an initialization for the next update. The results obtained using the aforementioned initialization are better than with zero initialization. It is worth mentioning that the convex upsampling module used in the original RAFT implementation \cite{RAFT} is not used in RAFT-StereoPIV or RAFT-StereoPIV-Small since no downsampling of the original input during feature extraction takes place. In addition, normalization is not included in the GRU updates, because it can be problematic for regression tasks like optical flow \cite{RAFT}.
Furthermore, by utilizing automatic mixed precision instead of single-precision in our PyTorch implementation, we realized a notable enhancement in computational efficiency. This approach led to a reduction in training time by approximately 30$\%$ streamlining our computational process.

\subsection{Results}
\subsubsection{Comparative Analysis on Problem Class 1}

Table \ref{tab:PC1_comaprison} presents a comparative analysis between state-of-the-art deep optical flow learning methodologies and our models, RAFT-StereoPIV and RAFT-StereoPIV-Small, on the unseen test dataset, Problem Class 1. The best values from each experiment are highlighted in bold. RAFT-StereoPIV and RAFT-StereoPIV-Small outperform all other methods across all flow instances. 

\begin{table*}[h!]
\begin{center}
    \begin{tabular}{llllll} 
    \multicolumn{5}{c}{State-of-the-art deep learning methods' averaged endpoint error for Problem Class 1} \\  
    \hline
Method & 
Cylinder & JHTDB-channel & DNS-turbulence & SQG \\
\hline
UnLiteFlowNet32-PIV  \cite{lagemann2021deep} & 
65.9 & 41.9 & 44.3 & 40.1 \\
UnLiteFlowNet-PIV \cite{zhang2020unsupervised} & 
7.9 & 14.5 & 22.5 & 21.6 \\
LightPIVNet  \cite{yu2021lightpivnet} & 
5.2 & 22.8 & 19.7 & 24.9 \\
RAFT32-PIV \cite{lagemann2022generalization} & 
2.1 & 3.6 & 6.8 & 5.9 \\
RAFT-StereoPIV-Small (Ours) & 
\textbf{1.3} & \textbf{1.3} & \textbf{3.4} & \textbf{1.8} \\
RAFT-StereoPIV (Ours) & 
$\mathbf{1.3}$ &  $\mathbf{1.3}$ & $\mathbf{3.3}$ & $\mathbf{1.9}$
\end{tabular}
\caption{Comparison between the state-of-the-art deep optical flow learning methods and our models RAFT-StereoPIV and RAFT-StereoPIV-Small on the unseen test dataset Problem Class 1. The unit of averaged endpoint error is normalized to pixels per 100 pixels. The best values from each experiment are marked in bold. RAFT-StereoPIV and RAFT-StereoPIV-Small achieve the best performance on all the flow cases.
}
\label{tab:PC1_comaprison}

\end{center}
\end{table*}





\subsubsection{Systematic Investigation of the Effects of Particle Image Parameters}
Our systematic analysis of synthetic PIV images showcases the RAFT-StereoPIV model's adaptability and robustness. We compare RAFT-StereoPIV with RAFT32-PIV \cite{lagemann2022generalization}. The study reveals the following results (see Table \ref{tab:SummarySynthetic}):

\begin{table*}[htb!]
\centering
{\small
\begin{tabular}{c|c|cccc}
    \multicolumn{6}{c}{Averaged endpoint error for synthetic images with varying particle image
conditions} \\  
\hline
\multirow{3}{*}{Parameter}  & \multirow{3}{*}{Value} &\multicolumn{4}{c}{Experiment}  \\ \cline{3-6} 
 &  & Cylinder & $\begin{array}{c}\text{JHTDB} \\ \text{Channel}\end{array}$ & $\begin{array}{c}\text{DNS} \\ \text{turbulence}\end{array}$ & SGQ \\
\hline
 & $0.5-1.25$ & 
$\mathbf{0.86}$ (1.01) & $\mathbf{0.77}$ (0.79) & $\mathbf{1.01}$ (1.02) & $\mathbf{0.84}$ (0.94) \\
& $1.25-2.5$ 
& 
$\mathbf{0.56}$ (0.64) & $\mathbf{0.62}$ (0.66) & $\mathbf{0.65}$ (0.79) & $\mathbf{0.77}$ (0.83) \\
$d_p$ &  $2.5-3.75$ 
&  
$\mathbf{0.60}$ (0.61) &  $\mathbf{0.63}$ (0.75) & $\mathbf{0.73}$ (0.83) &  $\mathbf{0.84}$ (0.86) \\
& $3.75-5$ 
& 
$\mathbf{0.61}$ (0.70) & $\mathbf{0.58}$ (0.66) & $\mathbf{0.77}$ (0.81) & $\mathbf{0.94}$ (0.97) \\
& $5-7$ & 
$\mathbf{0.73}$ (0.84) & $\mathbf{0.79}$ (0.91) & $\mathbf{1.02}$ (1.03) & $\mathbf{1.02}$ (1.12) \\
\hline
&  0.01 
& 
$\textbf{1.38}$ (1.95) & $\textbf{1.43}$ (2.14)  & $\textbf{2.19}$ (2.51) & $\textbf{1.47}$ (2.16) \\
$N_{p p p}$ &  0.05  
&  
$\mathbf{0.60}$ (0.61) &  $\mathbf{0.63}$ (0.75) & $\mathbf{0.73}$ (0.83) &  $\mathbf{0.84}$ (0.86) \\
&  0.1 
&  
0.44 ($\mathbf{0.32}$) &  0.38 ($\mathbf{0.33}$) & $\mathbf{0.41}$ (0.50) & 0.69 ($\mathbf{0.60}$) \\
\hline
&  4.3 
& 
0.47 ($\textbf{0.45}$) & $\textbf{0.46}$ (0.49) & $\textbf{0.52}$ (0.80) & $\textbf{0.62}$ (0.69) \\
SNR & 2.9 
&  
$\mathbf{0.60}$ (0.61) &  $\mathbf{0.63}$ (0.75) & $\mathbf{0.73}$ (0.83) &  $\mathbf{0.84}$ (0.86) \\
&  2.0 
& 
$\textbf{0.75}$ (0.86) & $\textbf{0.59}$ (0.73) & $\textbf{0.64}$ (0.75) & $\textbf{0.67}$ (0.78) \\
\hline 
\end{tabular}}
\caption{Summary of the performance of RAFT-StereoPIV and (RAFT32-PIV \cite{lagemann2022generalization}) on synthetic PIV images, specifically assessing the impact of particle diameter, particle density, and signal-to-noise ratio on averaged endpoint error (AEE) values. The best values from each experiment are marked in bold.}
\label{tab:SummarySynthetic}
\end{table*}

\begin{itemize}
    \item  The first five experiments focus on the effect of varying particle diameters ($d_p$), providing evidence of the RAFT-StereoPIV model's robustness to varying $d_p$ values. 
Contrasting with the RAFT32-PIV, which experiences a noticeable degradation in performance when particle diameters exceed double the size of the training distribution. The RAFT-StereoPIV model exhibits strong robustness to fluctuations in particle diameters, tolerating a range from 0.5 to 7 pixels, even though particle diameters larger than 5 pixels fall outside the domain distribution.

    \item The second set of experiments investigates the error sources related to varying particle densities ($N_{ppp}$). The results indicate that higher $N_{ppp}$ values correspond to lower AEE values for both models as more feature information becomes available.
 
     \item The final set of experiments examines the effect of varying signal-to-noise ratios (SNR), encompassing both changes in noise levels and peak intensities. For both models, the AEE's progression is similar, revealing that higher SNR values result in lower performance.

\end{itemize}
This thorough examination of synthetic PIV images with diverse parameters allows us to assess the RAFT-StereoPIV model's generalizability and robustness across various scenarios. RAFT-StereoPIV outperforms RAFT32-PIV baseline method and achieves state-of-the-art performance on all test cases.  

\subsubsection{Hyperparameters Analysis}
Our hyperparameters analysis for the RAFT-StereoPIV model yielded several significant findings. We discovered a trade-off between batch size and variance in gradients, with larger batch sizes delivering better performance, albeit limited by available GPU memory. An increase in dropout probability enhanced the model's performance but required an extended training duration for optimal generalization on unseen data. Among the various normalization methods tested, including batch, instance, group, and no normalization, instance normalization emerged as the most robust, despite group normalization initially showing superior performance during training and validation. Moreover, we found that using a training dataset of 20,000 samples yielded good results, but expanding the dataset further could still be beneficial by serving as a regularization technique and helping reduce overfitting. Lastly, the number of training epochs, indicating the frequency of passing the training dataset through the network, significantly influenced the model's convergence and optimization. These results provide valuable insights for optimizing the RAFT-StereoPIV model, aiding in our larger goal of democratizing access to large-scale deep learning models by leveraging community resources and exploring the hyperparameters space.

By fine-tuning RAFT-StereoPIV's hyperparameters, we managed to cut down the training time from roughly two days to 13.6 hours, resulting in a 73\,$\%$ reduction in training time when using 2 Quadro GV100 GPUs, thereby greatly enhancing the computational efficiency of the model. Most notably, our efforts led to a substantial decrease in error rates: a 68\,$\%$ reduction on the validation dataset, Problem Class 2, which contains 14,400 fluid flow test cases, and a 47\,$\%$ reduction on the unseen test dataset, Problem Class 1, with 10,000 fluid flow test cases. The detailed hyperparameters analysis is given in \ref{Appendix:HyperparametersTuning}.

\begin{figure*}[htbp]
    \centering
 \hspace{-8mm} 
 \includegraphics[width=\textwidth]{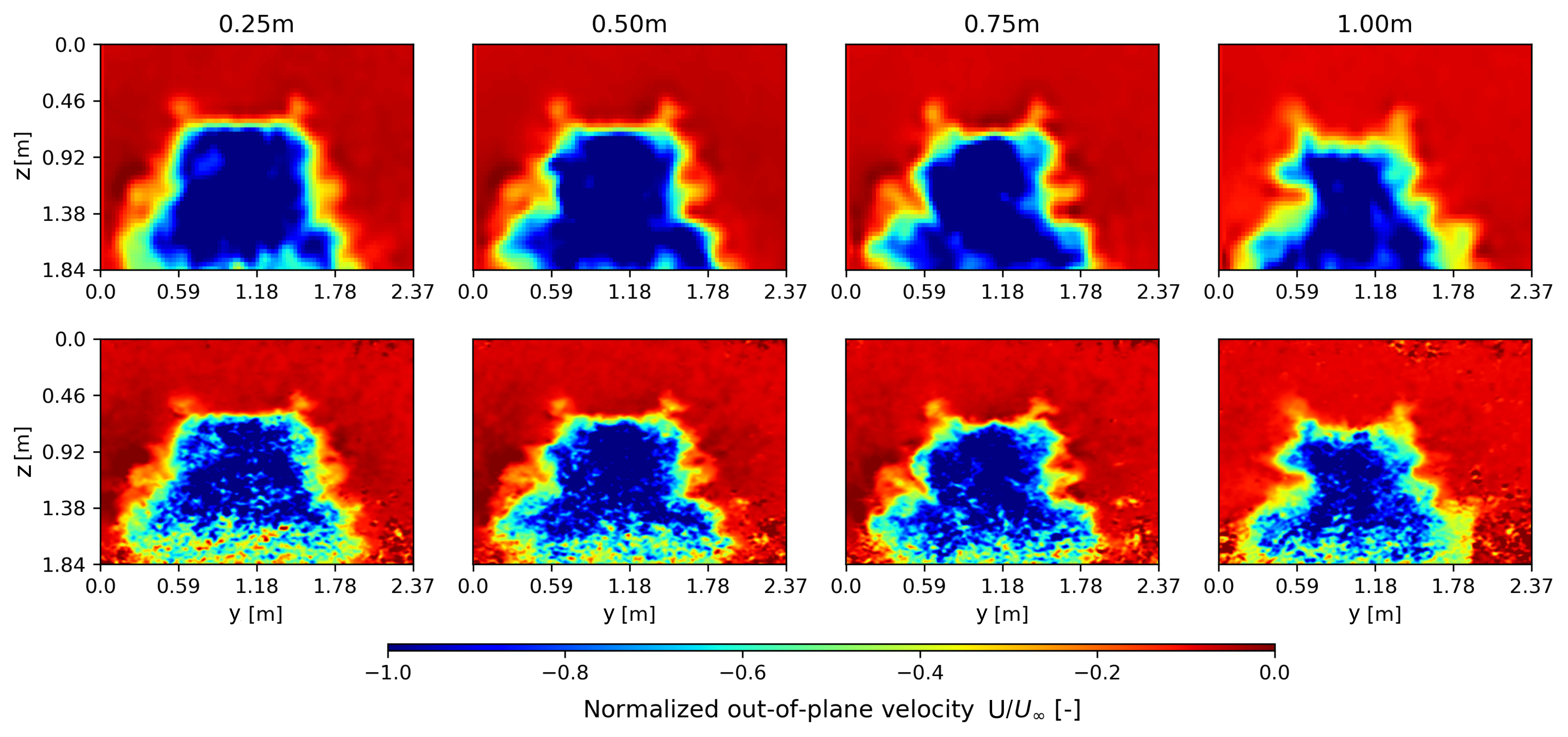}
    \caption{Ring of Fire experimental data -
    Development of the normalized out-of-plane streamwise velocity component in the wake of the car at $x$ = 0.25, 0.50, 0.75, and 1.00 m. (Top) we show the results obtained from LaVision DaVis 10.2 after postprocessing and 
    (bottom) - from the deep learning model RAFT-StereoPIV.
     RAFT-StereoPIV outputs dense per-pixel optical flow field, whereas the cross-correlation-based LaVision DaVis 10.2 outputs only sparse velocity field with 40 $\times$ 30 vectors.}
    \label{fig:ComparisonDl_DavisNew}
\end{figure*}
\subsubsection{Flow Field Estimation from Ring of Fire}
\begin{figure*}[htbp]
    \centering
    \includegraphics[width=0.9\textwidth]{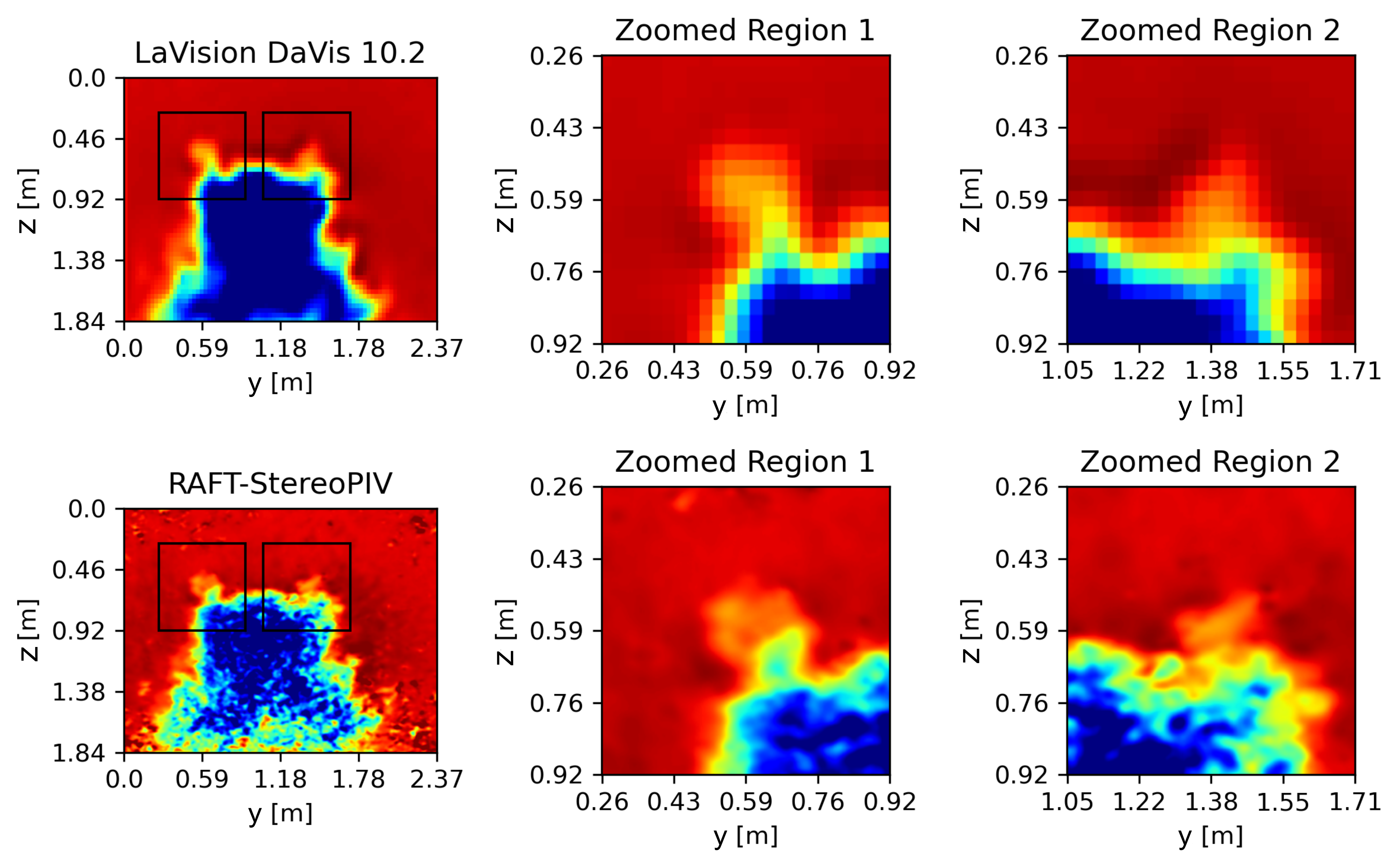}
    \caption{Spatial resolution comparison in the wake of the car from Ring of Fire  - Highlighting the difference in the spatial resolution between the sparse vector field output from the cross-correlation based LaVision DaVis 10.2 (top) and the dense per-pixel optical flow field output from RAFT-StereoPIV (bottom).
    }
    \label{fig:SpatialResolutionComparison}
\end{figure*}

We evaluate the performance of RAFT-StereoPIV on the challenging time-resolved experimental data from RoF. Since no ground-truth optical flow exists, we benchmark our results with the cross-correlation-based commercial software DaVis 10.2 from LaVision GmbH.

Reducing the size of the interrogation window is the simplest approach to improve the spatial resolution when using traditional methods, although doing so increases the uncertainty \cite{spoelstra2021uncertainty}. Therefore, given the camera resolution, a compromise between an image size large enough to capture the entire wake of the car and an interrogation window small enough to capture the small-scale structures must be found. At the same time, it is necessary to maintain an appropriate SNR to reduce the number of spurious velocity vectors as well as the uncertainty on the estimated drag.  For the flow field estimation using DaVis 10.2, we consider the following processing steps \cite{huttig2023automotive}, \cite{huettig_RoF}:
sliding sum of correlation function, followed by multi-pass and finally, universal outlier detection \cite{westerweel2005universal} for post-processing. Multi-pass comprises
\begin{itemize}
    \item a single pass with a 64$\times$64 interrogation window, 50\,$\%$ overlap, and square weighting
    \item three passes at a window size of 48$\times$48, 75\,$\%$ overlap and Gaussian weighting.
\end{itemize} 
Therefore, the smallest interrogation window that will be used for DaVis 10.2 is 48$\times$48 with 75\,$\%$ overlap. This results in a highly averaged flow field information. In comparison, the results obtained from our model RAFT-StereoPIV are dense per-pixel optical flow field information. 

Since the resolution of the images from most experimental data is much larger compared to the training data resolution ($32 \times 32 \, \mathrm{px}^2$), we implement a folding/unfolding scheme that generates cutouts of the input images. As a result, multiple forward passes are performed when processing large PIV images.


Overall, Figure \ref{fig:ComparisonDl_DavisNew} shows comparable flow structures at different positions in the wake of the car for both methods, namely DaVis 10.2 and RAFT-StereoPIV. However, sharp and more detailed flow field information
is achieved with the deep optical flow learning model RAFT-StereoPIV.
In contrast to the results from  DaVis 10.2  the data handling and the flow field estimation using deep learning, including processing the PIV images, can be completely automated and simultaneously done on multiple GPUs. Neither preprocessing nor postprocessing operations were applied to the images or the resultant data from RAFT-StereoPIV.





It is observable that the predictions derived from RAFT-StereoPIV encompass certain outliers. Employing an outlier detection, such as universal outlier detection \cite{westerweel2005universal}, has the potential to mitigate these outliers. However, the primary emphasis of this study is to present the unprocessed predictions generated by the deep learning model, as this approach facilitates the measurement of the instantaneous three-dimensional velocity field with high resolution and enables the visualization of fine turbulent flow structures.

Figure \ref{fig:SpatialResolutionComparison}. illustrates a spatial resolution comparison in the car's wake from RoF experiment. This comparison highlights the contrast between the sparse vector field output from the cross-correlation based LaVision DaVis 10.2 (shown at the top) and the detailed per-pixel optical flow field from RAFT-StereoPIV (shown at the bottom). Applying our trained network allowed us to maintain the spatial resolution of the original image sequence, a significant enhancement over traditional methods that typically produce flow fields at a resolution about 16 times lower.

\subsubsection{Flow Field Estimation from Wind Tunnel Measurements}

The PIV analysis of the flow field in the wake of the Ahmed body consists of several steps. For the reduction of background noise and reflections a directional minimum filter with a filter length of  $L=11\,$pixels was applied, leading to a better signal-to-noise ratio.  Consequently a multi-pass PIV-calculation was performed (\nth{1}\,pass: window size 96$\times$64, square weighting, 50\,\% overlap; \nth{2}\,to\,\nth{3}\,pass: window size 16$\times$16, Gaussian weighting, 75\,\% overlap), leading to a grid spacing of 0.165~mm. Post-processing included the universal outlier detection (\cite{westerweel2005universal}) within a region of 5$\times$5\, vectors, followed by linear interpolation to fill up empty spaced, and finally all 100 instantaneous vector fields were averaged.
Here, in contrast to RoF we apply RAFT-StereoPIV on the preprocessed double-frame particle images and not the raw data. 

\begin{figure*}[h!]
    \centering
    \includegraphics[width=0.7\textwidth]{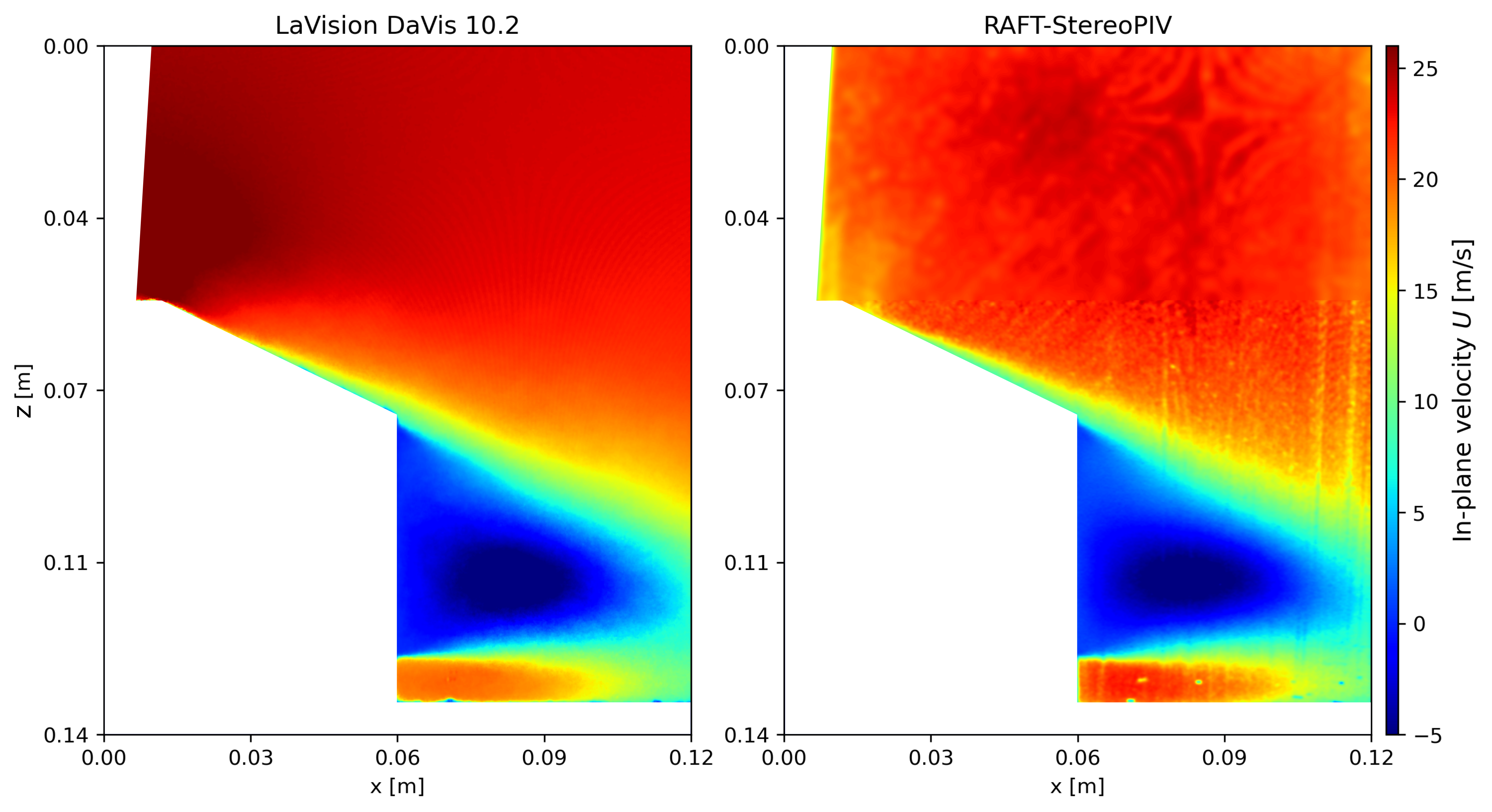}
    \caption{Wind Tunnel Measurements - In-plane mean velocity component $\overline{u}$  in the streamwise symmetry plane of the Ahmed body at $y$~=~0. (Left) The results obtained from LaVision DaVis 10.2 and (right) - from the deep learning model RAFT-StereoPIV. RAFT-StereoPIV outputs dense per-pixel optical flow field 3000 $\times$ 3519, whereas the cross-correlation-based LaVision DaVis 10.2 outputs only sparse velocity field with 750 $\times$ 880 vectors.
    }
    \label{fig:AhmedBodyComparison}
\end{figure*}

In summary, a good agreement between the flow structures obtained from DaVis 10.2 and RAFT-StereoPIV, as can be seen in Figure \ref{fig:AhmedBodyComparison}. This includes the recirculation zone in the wake as well as the separation point. 
However, due to the chosen angular stereo setup and remaining surface reflection, which could not be completely eliminated in the pre-processing step, there exist some discrepancies, such as outliers and artifacts.

\section{Limitations and Future Work}
\label{sec:sec5_FutureWork}

Deep learning in Particle Image Velocimetry shows great potential but faces challenges such as data dependency, lack of interpretability, and high computational demands. Future efforts in deep learning-based PIV should explore the development of models trained to integrate specific preprocessing filters. These would target noise reduction, outlier removal, or minimizing reflections, thereby enhancing data quality before analysis. 
Additionally, improving data augmentation, synthetic and wind tunnel data utilization, transfer learning for broader generalization, model interpretability, and computational efficiency remains crucial. Further advancements in real-time processing could also broaden the scope of deep learning applications in PIV.

\section{Conclusions}
By integrating advanced deep learning techniques with both Ring of Fire and wind tunnel measurements, we are able to achieve flow field visualization with high spatial resolution, a key element in aerodynamic testing and vehicle development. We achieved a new state-of-the-art on all benchmarks, in terms of both prediction accuracy and training time. Recurrent All-Pairs Field Transforms for Stereoscopic Particle Image Velocimetry (RAFT-StereoPIV) surpasses all existing supervised and unsupervised learning-based approaches in performance. 
In addition, it is the first flow estimation based on 3D reconstruction from 2D stereoscopic PIV data using deep learning. 
Our proposed method has the potential for real-time and on-site characterization of automotive aerodynamics under real driving conditions, as well as in controlled wind tunnel environments. As the vehicle passes through the laser sheet, as in the open-road setting Ring of Fire, RAFT-StereoPIV effectively visualizes the flow field in the vehicle's wake. This capability provides a detailed and dynamic insight into the aerodynamic interactions, crucial for comprehensive analysis.

We believe that this study helps pave the way forward to enable the development of more advanced measurement techniques for stereoscopic PIV in general and industrial aerodynamics in particular.

\subsubsection*{Author Contributions}
ME handled the development and implementation of the proposed method and testing on provided data.
ME, with MG, designed, trained, tuned, and evaluated 
the deep learning models.
ME proposed the 3D optical flow reconstruction module and started the effort with SH and TG to evaluate this idea. 
SH and TG were responsible for conducting the RoF experiment and the wind tunnel measurements of Ahmed Vehicle Model. TG, DC, and CB were responsible for research supervision. All co-authors meaningfully contributed to manuscript writing and review prior to submission.


\bibliography{bibliographie.bib}

\subsubsection*{Acknowledgements}
We thank Christian Lagemann for many fruitful discussions and valuable feedback. We also express our gratitude to Steffen Ropers for his assistance in training the deep learning models.

\subsubsection*{Disclaimer}
The results, opinions, and conclusions expressed in this publication are not necessarily those of Volkswagen Aktiengesellschaft.

\subsubsection*{Conflict of interest}
The authors declare that they have no conflict of 
interest.

\subsubsection*{Data Availability Statement}
This study used both open-source datasets and proprietary data from Volkswagen AG. The proprietary data from Volkswagen AG is confidential and not publicly available. Access was granted under a specific agreement and may be available upon request, subject to Volkswagen AG's approval and data sharing policies.

\renewcommand{\thefigure}{A\arabic{figure}}
\renewcommand{\thetable}{A\arabic{table}}
\setcounter{figure}{0} 
\setcounter{table}{0}  


\newpage
\onecolumn 
\appendix
\section{RAFT-StereoPIV Approach}

The pseudo-code for our approach, where we refer to the 4D description of the flow field as the 3D flow information estimated sequentially over time, is provided in Algorithm.~\ref{alg:4dflow}.

\begin{algorithm}[h!]
\floatname{algorithm}{Algorithm} 
\textbf{Input:} $n$ grayscale image-pairs  $\{ I^{t}$, $I^{t + \Delta t} \}$, calibration angles $\{\alpha_1,\alpha_2, \beta_1, \beta_2 \}$, magnification factor $M$ 
\vspace{3pt}
\newline
Use mapping function to map the input images:  $f(I^{t}$, $I^{t + \Delta t}) = (\tilde{I}^{t},\tilde{I}^{t + \Delta t} )$
\begin{algorithmic}[1]
\For{$i=1 . . n$} \Comment{iterate over $n$ number of image-pairs}
\For{$j=1 . . 2$}  \Comment{iterate over two cameras}
\vspace{3pt}
\State \parbox[t]{\dimexpr0.9\linewidth-\algorithmicindent}{Estimate optical flow $\Delta x_{i,j}$, $\Delta y_{i,j}$ using the RAFT-StereoPIV model}
\vspace{5pt}
\State $u_{i,j} \gets \frac{\Delta x_{i,j}}{M \cdot \Delta t}$ ; $v_{i,j} \gets \frac{\Delta y_{i,j}}{M \cdot \Delta t}$
\vspace{5pt}
\EndFor
\vspace{3pt}
\State \parbox[t]{\dimexpr0.9\linewidth-\algorithmicindent}{Reconstruct 3D velocity components from 2D velocities:}
\vspace{3pt}
\State $U_i \gets \frac{u_{i,1} \tan \alpha_2+u_{i,2} \tan \alpha_1}{\tan \alpha_2+\tan \alpha_1}$,
\State $V_i \gets \frac{v_{i,1} \tan \beta_2+v_{i,2} \tan \beta_1}{\tan \beta_2+\tan \beta_1}$,
\State $W_i \gets \frac{u_{i,1}-u_{i,2}}{\tan \alpha_2+\tan \alpha_1}$
\vspace{3pt}
\EndFor
\State Ensemble averaging over $n$ image-pairs to reduce the effect of noise and outliers:
\State $\overline{U} \gets \frac{1}{n} \sum_{i=1}^n U_{i}$ , $\overline{V} \gets \frac{1}{n} \sum_{i=1}^n V_{i}$ , $\overline{W} \gets \frac{1}{n} \sum_{i=1}^n W_{i}$
\end{algorithmic}
\caption{4D Flow Field Estimation Using Stereoscopic PIV and Deep Learning}
\label{alg:4dflow}

\end{algorithm}

\section{Hyperparameters Tuning}
\label{Appendix:HyperparametersTuning}

In this study, we endeavor to facilitate the training of large-scale deep learning models using community resources. The high computational demands required to train those models create substantial obstacles to reproducibility and further development, especially for researchers who have limited access to high-performance computing infrastructure.
By leveraging community resources and systematically exploring the hyperparameters space, our approach aims to democratize access to the state-of-the-art models, thereby fostering a more inclusive research environment and stimulating advancements in the field. In addition, we aim to enhance the overall performance, generalizability, and quality of results on real-world data. To achieve this, our study involves conducting large-scale analysis of hyperparameters to optimize the training process of the underlying optical flow network.

To better understand the importance of specific hyperparameters of RAFT-StereoPIV and the performance difference based on the chosen set of hyperparameters, we ran several experiments.  Results are shown in Table \ref{tab:hyperparams}. 
We report the AEE values for both the training and validation datasets, however, during the process of selecting the optimal hyperparameter values, only the AEE values obtained from the validation dataset are taken into account. 
Sections (A)-(F) of the table test exclusively a single hyperparameter, such as batch size, dropout probability $P_{drop}$, normalization method, initial learning rate $\alpha_{init}$, the reduce factor $\gamma$, the size of the training dataset $D$, and the number of training epochs $N_{epochs}$. On the other hand, section (G) tests different combinations of these hyperparameters, based on results obtained from the previous experiments.
\\
\\
\textbf{Batch size:} Smaller batch size means greater variance in the gradients and thus noisy updates, however, having larger batch sizes is mostly limited by the GPU memory available. Therefore, we tested different batch sizes \{4,16,32,64,128,192\}  to explore the trade-off between variance in the gradients and memory requirements when selecting the batch size. From (A), we observe, in accordance with expectations, larger batch sizes lead to better model performance. \\
\\
\textbf{Dropout:} Our findings indicate that the model's performance improves as the dropout probability increases. However, it is essential to note that an extended training duration is necessitated to achieve optimal generalization capabilities on unseen test data, allowing the model to effectively learn robust representations from the data. \\
\\ 
\textbf{Normalization:} Our experiments showed that group normalization demonstrated superior performance on the training and validation datasets. However, when applied to the unseen test dataset, the model performed poorly, which led us to opt for instance normalization. One possible explanation for the poor performance of group normalization on the test dataset could be that it is less capable of generalizing to data with distribution shifts or variations not present in the training and validation sets. By contrast, instance normalization is more robust to such discrepancies, making it a more suitable choice for this application. \\
\\
\textbf{Learning rate:} We tested various learning rate values drawn from the literature and closely monitored the learning curves to assess the impact on the training dynamics and final model performance. A learning rate of 0.001 proved to be the most effective, yielding the best balance between training speed and model accuracy. This value facilitated steady convergence without causing the training to stall or overshoot the minimum loss.
\\
\\
\textbf{Size of training dataset:} The fundamental objective of experiment (E) was to address the inquiry of determining the minimum training set size necessary for achieving satisfactory model performance. Our findings suggest that a training dataset of 20,000 samples can still yield very good results. 
\\
\\
\textbf{Number of epochs:} Finally, the number of training epochs $N_{epochs}$ refers to the number of times the training dataset is passed through the network during training, which affects the convergence and optimization of the model. While longer training durations can lead to overfitting, we determined that the optimal number of epochs for our training was 120. 
\\
%

\begin{table*}[h!]
\centering
{\small
\begin{tabular}{c|cccccccc|c|c|c}
\hline
Experiment & Batch Size & $P_{drop}$ & Norm & $\alpha_{init}$ & $\gamma$ & $D$ & $N_{epochs}$ & & $\begin{array}{c}\text { Train } \\
\text { AEE }\end{array}$  & $\begin{array}{c}\text { Val } \\
\text { AEE }\end{array}$ & $\begin{array}{c}\text { Time } \\
\text { hrs }\end{array}$ \\
\hline
Base & 4 & 0.0 & 'instance'  &  0.0001 & 0.2 &  104k &   80 &  & 0.1084 & 0.1111 &  46.3 \\
\hline
 & 16 &   &   &      &   &  & &   &  0.0644 & 0.0663 &  21.2\\
 & 32 &   &   &      &   &  & &   &  0.0555 & 0.0577 & 17.6 \\
(A) & 64 &   &     &   &   & &  &   &  0.0499 & 0.0517 & 15.5\\
 & 128 &   &   &     &   & &  &   &  0.0474 & 0.0495 & 14.2\\
 & \textbf{196} &   &   &  &  &   &   &   &  \textbf{0.0442} & \textbf{0.0459} &  \textbf{13.6}\\
 \hline
 & 196  & 0.1 &   &     &  & &   &   & 0.0379  & 0.0394 & 13.6\\
 &  196 & 0.2 &   &  &   &   &   &   & 0.0400 & 0.0415 & 13.6\\
 \multirow{3}{*}[4ex]{(B)}  & 196  & 0.3 &   &      &   &  & &   & 0.0375 & 0.0389 & 13.6\\
  & \textbf{196}  & \textbf{0.4} &   &   &   &   &   &   & \textbf{0.0348} & \textbf{0.0363} & \textbf{13.6}\\
 \hline
 & 196 &   &'batch' &     &   &   &  & &  diverged & diverged & 13.6\\
(C) &  \textbf{196}  &   &  \textbf{'group'}    &   &   & &  &   &  \textbf{0.0361}  &  \textbf{0.0374} & 13.6\\
 &  196  &   & 'none' &      &   & &  &   & 1.6160  & 1.6371 & 13.6\\
 \hline
 & 196  &   &      &  0.01  &  & &   &   & diverged &  diverged& 13.6\\
 &  \textbf{196} &   &      &  \textbf{0.001} & &   &   &   &  \textbf{0.0552} & \textbf{0.0568} & \textbf{13.6}\\
\multirow{3}{*}[4ex]{(D)}  &  196  &      &   &  0.00001 &  & &   &   & 0.1458 & 0.1512 & 13.6\\
 &  196 &   &   &      0.000001 &   &   & &  &   0.6235 & 0.6368 & 13.6\\
\hline
 & 196  & &   &     & &  20k &     & &  0.0349  &  0.0363 & 2.7\\
  & 196  & &   &    &  &  40k &      & & 0.0340 & 0.0354 & 5.2\\
\multirow{3}{*}[4ex]{(E)} &  \textbf{196}&  &     & &  & \textbf{60k}&   &    & \textbf{0.0332}& \textbf{0.0346} & \textbf{7.8}\\
 & 196  &  &   &    &  & 80k  &   &   & 0.0338 &  0.0352 & 10.5\\
\hline
 & 196  & &   &    & &   &  40 &   &  0.0327 &  0.0341 & 6.8\\
 (F) & \textbf{196}  & &   &  &   &   &  \textbf{120 } &   &  \textbf{0.0339} & \textbf{0.0354}  & \textbf{20.4}\\
 &  196 &  &     &  & &   & 160  &   & 0.0370 & 0.0384 & 27.2\\
 \hline
 & \textbf{196}  & \textbf{0.4} &   &    & \textbf{0.05} &   &  \textbf{100} &   &  \textbf{0.0333 }& \textbf{0.0348}  & \textbf{17.1}\\
 (G) & 196  & 0.4 &  & &   0.05 &   &  120 &   &  0.0339 &  0.0352 & 20.4\\
 &  196 &  0.5 &     & & 0.05 &   & 160  &   & 0.0343 & 0.0357 & 27.2\\
\hline
\end{tabular}}
\caption{Comprehensive results of the large-scale hyperparameter exploration experiments (RAFT-StereoPIV). Unlisted values are identical to those of the base model. The models were trained on two Quadro GV100 GPUs. The best values from each experiment are marked in bold.}
\label{tab:hyperparams}
\end{table*}

\end{document}